\documentclass[11pt,a4paper]{article}

\usepackage[english]{babel}
\usepackage[margin=1in]{geometry}
\usepackage{setspace}
\usepackage{microtype}

\usepackage{authblk}

\usepackage{amsmath}
\usepackage{amssymb}
\usepackage{amsfonts}
\usepackage{amsthm}
\usepackage{mathrsfs}
\usepackage{bm}
\usepackage{dsfont}
\usepackage{bbm}

\allowdisplaybreaks

\usepackage{graphicx}
\usepackage{float}
\usepackage{adjustbox}
\usepackage{booktabs}
\usepackage{makecell}
\usepackage{subcaption}
\usepackage{array}
\usepackage[flushleft]{threeparttable}
\usepackage{multirow}
\usepackage{longtable}
\usepackage{tabularx}
\usepackage{ltablex}

\usepackage{enumerate}
\usepackage{enumitem}

\setlist[enumerate]{label={\upshape(\roman*)}}

\usepackage[numbers]{natbib}
\usepackage{xcolor}
\usepackage[textsize=footnotesize]{todonotes}
\usepackage{times}

\usepackage[
    colorlinks=true,
    allcolors=blue
]{hyperref}

\theoremstyle{plain}

\newtheorem{assumption}{Assumption}
\newtheorem{theorem}{Theorem}
\newtheorem{lemma}{Lemma}
\newtheorem{proposition}{Proposition}

\theoremstyle{definition}

\newtheorem{definition}{Definition}

\theoremstyle{remark}

\newtheorem{remark}{Remark}

\newcommand{\rd}{\mathrm{d}}
\newcommand{\R}{\mathbb{R}}

\newcommand{\N}{\mathbb{N}}

\newcommand{\E}{\mathbb{E}}

\newcommand{\tl}{\tilde}

\newcommand{\sg}{\sigma}
\newcommand{\gm}{\gamma}
\newcommand{\Th}{\theta}
\newcommand{\lbd}{\lambda}
\newcommand{\ep}{\epsilon}

\newcommand{\lf}{\left}
\newcommand{\rt}{\right}

\newcommand{\cL}{\mathcal{L}}
\newcommand{\cF}{\mathcal{F}}

\newcommand{\ftf}{\lfloor t \rfloor}
\newcommand{\ctc}{\lceil t \rceil}

\title{
Non-asymptotic analysis of SGLD with
discontinuous stochastic gradients:
a new route to portfolio CVaR minimization
}

\author[1]{Yan Ding}
\author[2,3,4]{Sotirios Sabanis}
\author[1]{Ying Zhang}

\affil[1]{
Fintech Thrust,
The Hong Kong University of Science and
Technology (Guangzhou), China
}

\affil[2]{
School of Mathematics,
University of Edinburgh, UK
}

\affil[3]{
National Technical University of Athens,
Greece
}

\affil[4]{
Athena/Archimedes Research Centre,
Greece
}

\date{\today}

\begin{document}

\maketitle

\begingroup
\renewcommand{\thefootnote}{}
\footnotetext{
Financial support by the Guangzhou-HKUST(GZ) Joint Funding Program (No. 2025A03J3322) is gratefully acknowledged. Sotirios Sabanis was supported by Innovate UK [grant number 10081810] and was partially supported by project MIS 5154714 of the National Recovery and Resilience Plan Greece 2.0 funded by the European Union through the NextGenerationEU Program.
}
\endgroup

\begin{abstract}

Conditional-Value-at-Risk (CVaR) optimization is a
fundamental tool in portfolio selection, yet its
nonsmooth loss structure poses a challenge for the
existing theory of stochastic gradient methods
including associated Langevin algorithms.
In particular, the positive-part loss and the
softmax portfolio parametrization lead to stochastic
gradients that need not be globally Lipschitz
continuous, but rather discontinuous, where
available non-convex stochastic gradient Langevin
dynamics (SGLD) guarantees often exhibit an
unfavorable exponential dependence on the
problem dimension.

In this article, we develop a non-asymptotic
analysis of SGLD for a regularized CVaR portfolio
optimization problem.
We exploit the geometric structure of the CVaR
objective to establish linear growth,
convexity-at-infinity, and one-sided Lipschitz
properties of the stochastic gradient without
imposing density assumptions on the asset losses
that are typical in previous works.
Crucially, the constants appearing in the
aforementioned structural conditions are shown
to be independent of the number of assets.
Moreover, we establish a convergence bound in
Wasserstein-2 distance with rate
$\mathcal{O}(\lbd^{1/4})$ and a finite-time
guarantee for the expected excess risk.
Our analysis, together with the
dimension-independent constants in the structural
conditions, leads to an excess risk bound that
depends linearly on the number of assets.
Numerical experiments with simulated Gaussian
portfolios exhibit the predicted dimension
dependence.
Computational comparisons with alternative
stochastic optimization methods further
illustrate the practical efficiency of SGLD
for the optimization problem under consideration.
Finally, we extend the framework to a broader
class of non-convex stochastic optimization
problems with discontinuous stochastic gradients.

\end{abstract}


\section{Introduction}
Tail losses are a central concern in portfolio risk management, since losses occurring under adverse market conditions can have consequences far beyond those suggested by typical market fluctuations. Conditional-Value-at-Risk (CVaR) provides a natural criterion for controlling such risk, unlike Value-at-Risk (VaR), which specifies a loss threshold at a prescribed confidence level, CVaR also accounts for the severity of losses beyond that threshold. Although CVaR admits an explicit variational representation~\cite{MR1835507}, optimizing it from data leads to a challenging stochastic optimization problem. In particular, when the parameterized portfolio allocation and the associated VaR level are optimized jointly, the resulting problem exhibits both nonsmoothness and nonconvexity, placing it outside the standard globally smooth framework of stochastic optimization.

More precisely, the variational representation of CVaR introduced in~\cite{MR1835507} allows the portfolio allocation and the associated VaR level to be optimized simultaneously. Let $X^1,\ldots,X^N$ denote the asset losses and
$a=(a_1,\ldots,a_N)$ be the portfolio allocation. For a confidence level $\bar{q} \in (0,1)$, the resulting CVaR portfolio optimization problem is to minimize
\begin{equation}\label{CVaR1}
\eta+
\frac{1}{1-\bar q}
\E\left[
\left(
\sum_{i=1}^{N}a_iX^i-\eta
\right)_{+}
\right]
\end{equation}
over $\eta\in\mathbb R$ and $a\in A_N$, where $A_N:=\{a\in[0,\infty)^N:\sum_{i=1}^N a_i=1\}.$
A solution $(\eta^\ast,a^\ast)$ therefore yields both an optimal portfolio allocation $a^\ast$
minimizing CVaR and the corresponding VaR level $\eta^\ast$.

To obtain an unconstrained formulation, we parameterize the portfolio weights using the softmax mapping $g_i(w):=e^{w_i}/\sum_{j=1}^N e^{w_j}$ for all $i=1,\ldots,N,$ and consider the following regularized optimization problem
\begin{equation}\label{CVaR2}
\min_{\Th \in \mathbb R^{N+1} }V(\theta)
=
\min_{\Th \in \mathbb R^{N+1} } \lf(\E\left[
\eta+
\frac{1}{1-\bar q}
\left(
\sum_{i=1}^{N}g_i(w)X^{i}-\eta
\right)_{+}
\right]
+
\gamma|\theta|^{2}\rt),
\end{equation}
where $\theta=(\eta,w_1,\ldots,w_N)$ and $\gamma>0$ is a regularization parameter. Since the objective in \eqref{CVaR2} is defined through an expectation, stochastic approximation methods provide a natural framework when the distribution of the asset losses is accessible only through samples. However, this formulation exhibits several structural difficulties. In particular, the max function leads to stochastic gradients involving discontinuous indicator functions, and the nonlinear parametrization of the portfolio weights generally renders the objective nonconvex in the unconstrained variable $w$. Moreover, the gradient of the expected objective need not be globally Lipschitz continuous.
Consequently, standard stochastic gradient analyses based on globally smoothness cannot be applied directly.

To seek global optimization guarantees in this nonconvex setting, we study the stochastic gradient Langevin dynamics (SGLD) algorithm
\begin{equation}\label{introsgld}
\theta_{0}^{\lambda}=\theta_{0},
\qquad
\theta_{n+1}^{\lambda}
=
\theta_{n}^{\lambda}
-
\lambda H(\theta_{n}^{\lambda},X_{n+1})
+
\sqrt{2\beta^{-1}\lambda}\,\xi_{n+1},
\qquad n\in\mathbb N_{0},
\end{equation}
where $\Th_0$ is an $\R^{N+1}$-valued random variable, $\lambda>0$ is the stepsize, $H:\R^{N+1}\times \R^{N} \rightarrow \R^{N+1}$ is a measurable function satisfying $\nabla V(\Th) = \E[H(\Th, X_0)]$ for each $\Th \in \R^{N+1}$, $(X_n)_{n\in\mathbb N_0}$ is a sequence of i.i.d. $\R^N$-valued asset-loss vectors, $\beta>0$ is the inverse temperature parameter, and $(\xi_n)_{n\in\mathbb N}$ is a sequence of independent standard Gaussian random vectors. The SGLD algorithm \eqref{introsgld} can be viewed as an Euler discretization of the Langevin SDE:
\begin{equation}\label{introsde}
Z_0 = \Th_0, \qquad dZ_t=-\nabla V(Z_t)\,dt+\sqrt{2\beta^{-1}}\,dB_t,
\end{equation}
where $(B_t)_{t\geq0}$ represents a standard Brownian motion. Under appropriate conditions, the Langevin SDE~\eqref{introsde} admits the Gibbs measure $
\pi_{\beta}(d\theta)
\propto
e^{-\beta V(\theta)}\,d\theta$
as its invariant distribution. For sufficiently large $\beta$, the measure $\pi_\beta$ concentrates around the global minimizers of $V$, see, e.g., \cite{MR602391}. Thus, the minimization problem can be approached by approximately sampling from $\pi_\beta$. The central question considered here is therefore whether SGLD~\eqref{introsgld} can be analyzed for the nonconvex CVaR objective~\eqref{CVaR2} in the presence of discontinuous stochastic gradients and without assuming global Lipschitz continuity of $\nabla V$, while retaining quantitative bounds for expected excess risk that remain effective as the number of assets increases.

\subsection{Literature Review}
An early stochastic approximation approach was proposed by \cite{Pages} for simultaneously estimating the VaR and CVaR of a prescribed loss distribution. The method is based on the convex objective~\eqref{CVaR1} and the convergence analysis requires continuity of the gradient of the expected objective. Importantly, the portfolio allocation is fixed in the setting, so the resulting optimization is only with respect to the scalar VaR parameter and remains convex. The problem therefore concerns the estimation of VaR and CVaR for a given loss distribution, rather than the joint optimization of the portfolio allocation and VaR parameter considered in \eqref{CVaR2}.

More recently, multilevel stochastic approximation methods have been developed for settings in which the underlying loss is itself a conditional expectation and can only be approximated through nested Monte Carlo simulation; see \cite{Multilevel1,Multilevel2,Multilevel3}. These methods substantially reduce the computational cost of VaR and Expected Shortfall estimation in nested simulation problems. However, these multilevel studies primarily estimate the VaR or expected shortfall of a fixed portfolio and do not optimize the asset allocation. Moreover, their convergence analyses also impose regularity assumptions on the densities of the approximate losses, including uniform boundedness and Lipschitz continuity. Since the derivative of the one-dimensional objective~\eqref{CVaR1} is determined by the distribution function of the loss, uniform boundedness of the density implies Lipschitz continuity of the gradient of the expected objective, while Lipschitz continuity of the density imposes an additional second-order regularity condition. Furthermore, the expected objectives considered in these works remain convex.

Meanwhile, proximal methods provide an alternative framework for nonsmooth and nonconvex stochastic optimization. In \cite{Proximalstochastic}, a general stochastic model-based approach was developed for weakly convex objectives that may be both nonsmooth and nonconvex. The framework includes the proximal stochastic subgradient method and establishes complexity guarantees for finding approximate stationary points through the gradient of the Moreau envelope. Although CVaR minimization is included among the applications considered there, the theoretical guarantees concern approximate stationarity rather than global optimality, and the unconstrained portfolio formulation~\eqref{CVaR2} is not analyzed explicitly. From a sampling perspective, \cite{ProximalMCMC} studied a proximal stochastic gradient Langevin algorithm (PSGLA) for non-log-concave distributions with composite potentials $V=f+g$, combining an explicit gradient step for a smooth component with a proximal step for a nonsmooth weakly convex component and a Langevin perturbation. Non-asymptotic convergence guarantees are established for this proximal Langevin scheme in the nonconvex setting.

Non-asymptotic SGLD theory for nonconvex optimization has also been developed under progressively weaker regularity conditions. \cite{MR4313846} established uniform Wasserstein-1 convergence under a global Lipschitz condition on the stochastic gradient, while allowing dependent data streams. Subsequently, \cite{MR4534450} relaxed this requirement to local Lipschitz continuity and obtained convergence rates in both Wasserstein-1 and Wasserstein-2, together with an expected excess-risk bound. Nevertheless, both frameworks still require Lipschitz-type regularity of the stochastic gradient and therefore do not directly accommodate the discontinuous stochastic gradients arising in the CVaR formulation \eqref{CVaR2}.

Building on \cite{MR3945579}, where stochastic approximation with discontinuous updates was studied, the work most closely related to ours is the fully data-driven SGLD approach of \cite{sabanis2020fully}, which considers the unconstrained optimization problem \eqref{CVaR2}. In this setting, the regularized objective is generally nonconvex and the stochastic gradient is allowed to be discontinuous. Non-asymptotic Wasserstein convergence and expected excess risk estimates for SGLD are established under a continuity-in-average condition on the stochastic gradient. This condition, however, implies global Lipschitz continuity of the gradient of the expected objective. In addition, several constants appearing in the resulting convergence and excess risk bounds depend exponentially on the optimization dimension $d=N+1$, which may limit the effectiveness of these bounds as the number of assets increases.

A more recent development is the tamed stochastic gradient Hamiltonian Monte Carlo (tSGHMC) method proposed in \cite{wang2026tamed}, which addresses stochastic optimization with superlinearly growing and discontinuous stochastic gradients. They established non-asymptotic Wasserstein convergence and expected excess risk bounds under polynomial continuity-in-average and data-dependent strong convexity conditions, and further illustrated the method through a CVaR risk-allocation problem with a softmax parametrization. Their general excess risk bound exhibits polynomial dependence on the optimization dimension. When specialized to the parameter regime used in the CVaR application, this dependence is of order $(N+1)^{9/2}$. These results leave open the question of whether one can establish non-asymptotic guarantees for the nonconvex CVaR portfolio problem \eqref{CVaR2} with discontinuous stochastic gradients, without assuming global Lipschitz continuity of $\nabla V$, and with favorable dependence on the number of assets.

The present work addresses this problem by exploiting the specific geometric structure of the CVaR objective. In particular, we show that the stochastic gradient of \eqref{CVaR2} satisfies a linear growth condition, a convexity-at-infinity condition, and a one-sided Lipschitz condition, without imposing any density assumptions on the asset losses. Importantly, the constants entering these structural conditions can be chosen independently of the number of assets. These properties allow us to establish non-asymptotic Wasserstein convergence for SGLD under only second-moment assumptions on the initial condition and the asset losses. In particular, we obtain a Wasserstein-2 convergence rate of order $\mathcal{O}(\lbd^{1/4})$, with the relevant contraction and discretization constants growing only as $\mathcal{O}(\sqrt{N})$. This leads to an expected excess risk estimate whose dependence on $N$ is essentially linear.

To complement these theoretical comparisons, we also benchmark SGLD numerically against tSGHMC~\cite{wang2026tamed}, the proximal stochastic subgradient method~\cite{Proximalstochastic}, and PSGLA~\cite{ProximalMCMC} on the same CVaR portfolio optimization problem. While all four methods attain the prescribed expected excess risk accuracy, SGLD requires the lowest wall-clock time, followed by tSGHMC, whereas the two proximal methods incur substantially higher computational costs due to the numerical proximal updates used in our implementations.

The main contributions of this paper are summarized as follows. 
\begin{enumerate}
\item  We establish non-asymptotic convergence estimates for the SGLD iterates in Wasserstein-$2$ distances, see Theorem \ref{thm1}. More precisely, we obtain
\[
  W_2(\cL(\Th^{\lbd}_n),\pi_\beta) \leq 2C_5 \lbd^{1/4} +eC_6 e^{-n\lbd/T_0},
\]
where the discretization constant satisfies $C_5=\mathcal{O}(\sqrt{N})$ and the contraction constant $C_6 = \mathcal{O}(\sqrt{N})$. This improves the dimension dependence that appears in existing non-convex SGLD estimates.

\item Using the Wasserstein-$2$ convergence result, we derive a non-asymptotic upper bound for the expected excess risk 
\begin{equation}\label{introEER}
\E[V(\theta_n^\lambda)]-\inf_{\theta\in\mathbb R^{N+1}}V(\theta),
\end{equation}
see Theorem \ref{thm2}. We show that the upper estimate of \eqref{introEER} exhibits a linear dependence on the number of assets $N$. Numerical experiments further indicate an approximately linear growth of the expected excess risk with $N$, consistent with the theoretical result.

\item We extend the analysis beyond the particular CVaR problem to a general non-convex stochastic optimization problem
\[
\min_{\theta\in\mathbb R^{d}}
U(\theta),
\qquad
U(\theta)=\E[f(\theta,X)],
\]
whose stochastic gradient is allowed to be discontinuous. Under local versions of linear growth, convexity-at-infinity, and one-sided Lipschitz conditions, we establish analogous Wasserstein convergence and expected excess risk bounds for SGLD, see Theorems \ref{thm3} and \ref{thm4}. Thus, the CVaR portfolio problem therefore serves both as an important financial application and as a motivating example for a broader class of non-smooth stochastic optimization problems which extends the current state of the art.
\end{enumerate}

The remainder of the paper is organized as follows. In Section~\ref{Section2}, we introduce the CVaR problem under consideration and the associated SGLD algorithm. In Section~\ref{Section3}, we state the assumption and present the main convergence and expected excess risk results. Section~\ref{Section4} contains the numerical experiments for low-dimensional and high-dimensional Gaussian portfolios, together with the computational comparison with alternative algorithms. In Section~\ref{Section5}, we extend the analysis to a general class of non-convex stochastic optimization problems. The auxiliary estimates and proofs of the main results are provided in the appendices.

\subsection{Notation}We conclude this section by introducing some basic notations. For a positive real number $a$, we denote by $\lfloor a \rfloor$ its integer part, and define $\lceil a \rceil =\lfloor a \rfloor +1$. Let $\N_0:= \N \cup\{0\}$. Let $d,m\in \N$. For $x \in \R^d$, we write $|x|$ for its Euclidean norm and $x^{i}$ its $i$-th element. For $A \in \R^{d\times m}$, we denote by $A^{\top}$ its transpose. For $x,y \in \R^{d}$, we denote by $\langle x,y\rangle$ their Euclidean inner product. For any $\R^{d}$-valued random variable $X$, its law on $\mathcal{B}(\R^d)$, i.e., the Borel sigma-algebra of $\R^d$, is denoted by $\cL(X)$. Moreover, $\mathcal{P}(\R^d)$ denotes the set of all probability measures on $\mathcal{B}(\R^d)$, $C^2(\R^d)$ denotes the space of all twice continuously differentiable functions on $\R^d$, and $C_b^1(\R^{d})$ denotes the space of continuously differentiable functions on $\R^{d}$ that are bounded and have bounded first derivatives.

For $p\geq 1$, we denote by $\mathcal{P}p(\R^d)$ the set of all probability measures on $\R^d$ with finite $p$-th moment. For any $\gm,\nu\in\mathcal{P}p(\R^d)$, the Wasserstein distance of order $p$ is defined as
\[
W_p(\gm,\nu)
:=
\left(
\inf_{\zeta\in\Pi(\gm,\nu)}
\int_{\mathbb{R}^d\times\mathbb{R}^d} |x-y|^p\,\zeta(dx,dy)
\right)^{1/p},
\]
where $\Pi(\gm,\nu)$ denotes the set of all couplings of $\gm$ and $\nu$.

\section{CVaR of a portfolio of $N$ assets}\label{CVaR}\label{Section2}
We consider the problem of computing VaR and CVaR for a given portfolio of $N$ assets. This problem was discussed in \cite{MR1835507}, where it was formulated as a constrained convex optimization problem. In this paper, however, by applying an softmax parametrization to the asset weights, we reformulate the problem as the following \textit{unconstrained, non-convex} optimization problem. Then, by solving the problem, both the optimal weights and the corresponding VaR level can be obtained. 

The non-convex minimization problem under consideration is given below:
\begin{equation}\label{CVaRproblem}
\min_{ \Th \in \R^{N+1}} V(\Th)=\min_{\Th \in \R^{N+1}}\lf(\E\lf[\frac{1}{1-\bar{q}}\lf(\sum_{i=1}^N g_i(w)X^i-\eta\rt)_{+}+\eta\rt]+\gm|\Th|^2\rt),
\end{equation}
where $\bar{q}\in(0,1)$, $N \in [2,\infty) \cap \N$ is the number of assets in the portfolio, $\Th = (\eta, w)=(\eta,w_1,\ldots,w_N)\in \R^{N+1}$, $X^i$ denotes the loss of the $i$-th asset for $i=1,\ldots,N$, $g_i:\R^N \rightarrow \R$ denotes the weight of the $i$-th asset with
\begin{equation}\label{g_i(w)}
g_i(w):=\frac{e^{w_i}}{\sum_{j=1}^N e^{w_j}}\in(0,1), \;\;i=1,\ldots,N,
\end{equation}
and $\gm>0$ is the regularization parameter. Therefore, solving \eqref{CVaRproblem} yields not only the VaR level, but also the optimal portfolio weights that minimize the CVaR.

To solve \eqref{CVaRproblem}, we consider the SGLD algorithm given by
\begin{equation}\label{SGLD}
\Th_0^\lbd=\Th_0,\quad \Th_{n+1}^\lbd=\Th^\lbd_{n}-\lbd H(\Th^\lbd_{n},X_{n+1})+\sqrt{2\beta^{-1}\lbd}\xi_{n+1},\quad n\in\N_0,
\end{equation}
where $\Th_0$ is an $\R^{N+1}$-valued random variable, $\lbd>0$ is the step size, and $\beta>0$ is the inverse temperature parameter. The sequence $\lf(X_n\rt)_{n \in \N_0} = (X_n^1,\ldots,X_n^N)_{n \in \N_0}$ is assumed to be i.i.d. with common law $\cL(X)$, $\lf(\xi_n\rt)_{n \in \N}$ is a sequence of independent standard Gaussian random vectors in $\R^{N+1}$, and $H:\R^{N+1}\times \R^N \rightarrow\R^{N+1}$ is the stochastic gradient of $V$ with its explicit expression given by
\[
H(\Th,x):=(H_\eta(\Th,x),H_{w_1}(\Th,x),\ldots,H_{w_N}(\Th,x))^\top,
\]
where $H_\eta:\R^{N+1}\times \R^N \rightarrow \R$ and $H_{w_j}:\R^{N+1}\times \R^N \rightarrow \R$, $j=1,\ldots,N$ are defined by
\begin{equation}\label{Hetawj}
\begin{aligned}
    H_\eta(\Th,x)&:=1-\frac{1}{1-\bar{q}}\mathbbm{1}_{\{G(w,x)\geq\eta\}}+2\gm\eta,\\
     H_{w_j}(\Th,x)&:=\frac{1}{1-\bar{q}}\hat{g}_{w_j}(w,x)\mathbbm{1}_{\{G(w,x)\geq\eta\}}+2\gm w_j,
\end{aligned}
\end{equation}
with
\[
G(w,x):=\sum_{i=1}^Ng_i(w)x^i, \hspace{10mm} \hat{g}_{w_j}(w,x):=\frac{\partial G(w,x)}{\partial w_j}.
\]
For every $n\in \N_0$, let $\mathcal{G}_n:=\sg(X_k,k\leq n,k\in \N_0)$ and define $\mathcal{G}_{\infty} :=\sg(\cup_{n\in\N_0}\mathcal{G}_n)$. Throughout the paper, we assume that $\Th_0$, $\mathcal{G}_{\infty}$, and $\lf(\xi_n\rt)_{n\in \N}$ are independent.

Moreover, we denote by $h:=\nabla V$, and, for a fixed $\beta>0$ that is sufficiently large, we define the measure
\begin{equation}\label{gibbs}
\pi_{\beta}(A):=\frac{\int_{A} e^{-\beta V(\Th)}\rd \Th}{\int_{\R^{N+1}} e^{-\beta V(\Th)}\rd \Th},\qquad A\in \mathcal{B}(\R^{N+1}),
\end{equation}
provided that $\int_{\R^{N+1}} e^{-\beta V(\Th)}\rd \Th < \infty$. We note that minimizing $V$ in \eqref{CVaRproblem} is equivalent to sampling from $\pi_\beta$.

\begin{remark}
    For each $\Th \in \R^{N+1}$, $H$ defined in \eqref{Hetawj} satisfies $h(\Th) = \E[H(\Th,X_0)]$.
\end{remark}

\section{Assumptions and main results}\label{Section3}
In this section, we state the standing assumption and present the main convergence
result for the SGLD algorithm \eqref{SGLD}, see Theorem \ref{thm1}. Moreover, we derive a non-asymptotic error bound for the expected excess risk of the optimization problem \eqref{CVaRproblem}, namely $ \E[V(\Th_n^{\lbd})]-\inf_{\Th \in \R^{N+1}}V(\Th)$, in Theorem \ref{thm2}. The proofs are postponed to Appendix \ref{Proofs of main results}.

\begin{assumption}\label{A1}
	Assume that the initial value $\Th_0$ satisfies $\E|\Th_0|^2 < \infty$. The i.i.d. process $(X_n)_{n\in\N_0}$ satisfies $\E|X_0|^2 <\infty$.
\end{assumption}

\begin{remark}
Under Assumption~\ref{A1}, $H$ defined in
\eqref{Hetawj} satisfies a growth condition, a convexity-at-infinity
condition, and a one-sided Lipschitz condition, see
Lemmas~\ref{A2}, \ref{A3}, and \ref{A4}. Note that these conditions correspond to \cite[Assumptions~1, 2 and~4]{ULA}, respectively, under which non-asymptotic convergence guarantees are established for the unadjusted
Langevin algorithm (ULA) in Wasserstein distances. In addition, the conditions required for the Wasserstein
contraction result established in \cite{lytras2026tamed} are also satisfied by Lemmas~\ref{A2}, \ref{A3}, and \ref{A4}. Consequently, by adapting the convergence framework and arguments developed in \cite{ULA} for ULA to the SGLD algorithm~\eqref{SGLD} and combining them with the contraction estimate in \cite[Proposition~5]{lytras2026tamed}, we obtain the following main convergence result.
\end{remark}

\begin{theorem}\label{thm1}
  Let Assumption $\ref{A1}$ hold and set $\lbd_0=\min\{1,1/\gm,\gm/L\}$. Then, for any $\beta>0$, there exists constants $C_5$, $C_6$, and $T_0>0$ such that for any $\lbd \in(0,\lbd_0)$ and $n \in \N_0$, the SGLD algorithm \eqref{SGLD} satisfies,
  \[
  W_2(\cL(\Th^{\lbd}_n),\pi_\beta) \leq 2C_5 \lbd^{1/4} +eC_6 e^{-n\lbd/T_0},
  \]
 where $T_0$ is independent of the dimension $N$, $C_5=\mathcal{O}(\sqrt{N})$, and $C_6 = \mathcal{O}(\sqrt{N})$. The explicit expressions of $C_5$, $C_6$ and $T_0$ are given in Table \ref{table1}.
\end{theorem}

\begin{remark}
For $\ep>0$, the estimate in Theorem~\ref{thm1} implies that $W_2(\cL(\Th_n^\lbd),\pi_\beta)\leq \ep$ whenever $\lbd$ and $n$ are chosen such that
\[
\lbd \in(0,\lbd_0),\quad 2C_5 \lbd^{1/4}\leq \frac{\ep}{2},\quad
eC_6 e^{-n\lbd/T_0} \leq \frac{\ep}{2}.
\] 
Equivalently, it suffices to take
\[
\lbd\in
\lf(
0,
\min\lf\{
\lbd_0,
\frac{\ep^4}{256C_5^4}
\rt\}
\rt)
\quad
\text{and}
\quad
n\lbd
\geq
T_0\log\left(
\frac{2eC_6}{\ep}
\right).
\]
\end{remark}

By using Theorem \ref{thm1}, one can obtain the following expected excess risk bound for the CVaR objective \eqref{CVaRproblem}.

\begin{theorem}\label{thm2}
	 Let Assumption $\ref{A1}$ hold and set $\lbd_0=\min\{1,1/\gm,\gm/L\}$. Then, for any $\beta>\max\{4/\gm, 1/M\}$ with $M = 3\sqrt{L}/2+\sqrt{m}+\sqrt{2bL/\gm}$, there exist constants $C_V$, $C_5$, $C_6$, and $T_0 >0$ such that for any $n \in N_0$ and $\lbd \in(0,\lbd_0)$, the following bound holds:
	 \begin{align*}
	 \E[V(\Th_n^{\lbd})]-\inf_{\Th \in \R^{N+1}}V(\Th)&\leq  C_{V}\lf(2C_5 \lbd^{1/4} +eC_6 e^{-n\lbd/T_0} \rt) +\frac{2}{\beta}
	 -
	 \frac{1}{\beta}\log\lf(\frac{2\pi^{(N+1)/2}}{(N+1)\Gamma((N+1)/2)}\rt)\\
	 &\quad +	\frac{N+1}{2\beta}
	 \log\left(
	 \frac{4\pi e(b+(N+1)/\beta)\beta^2M^2}{\gm (N+1)}
	 \right),
	 \end{align*}
  where $T_0$ is independent of $N$, $C_V= \mathcal{O}(\sqrt{N})$, $C_5= \mathcal{O}(\sqrt{N})$, and $C_6= \mathcal{O}(\sqrt{N})$. The explicit expressions of $C_V$, $C_5$, $C_6$ and $T_0$ are given in Table \ref{table1}.	
\end{theorem}

\begin{remark}
For any $\ep>0$, by the estimate in Theorem \ref{thm2}, if we choose $\lbd$, $n$, and $\beta$ so that $2C_{V}C_5\lbd^{1/4}  \leq \ep/3$, $eC_{V}C_6e^{-n\lbd/T_0} \leq \ep/3$, and
\[
\frac{2}{\beta}
	 -
	 \frac{1}{\beta}\log\lf(\frac{2\pi^{(N+1)/2}}{(N+1)\Gamma((N+1)/2)}\rt)+\frac{N+1}{2\beta}
	 \log\left(
	 \frac{4\pi e(b+(N+1)/\beta)\beta^2M^2}{\gm (N+1)}
	 \right) \leq \frac{\ep}{3},
\]
then we have $\E[V(\Th_n^{\lbd})]-\inf_{\Th \in \R^{N+1}}V(\Th)\leq \ep$. This yields
\[
	\lbd \in\lf(0, \min\bigg\{\lbd_0,
\frac{\ep^4}{(6C_VC_5)^4}
	\bigg\}\rt),\qquad  	n\lbd \geq T_0
	\log\bigg(
	\frac{3eC_VC_6}{\ep}
	\bigg),
	\]
	and 
	\[
	\beta\geq \max\lf\{\frac{4}{\gm},\frac{1}{M},\frac{6C_{\beta_0}}{\ep},e,\frac{12(N+1)\log\big(12(N+1)/\ep\big)}{\ep} \rt\},
	\]
	where  
    \[
     C_{\beta_0}=2-\log\left(\frac{2\pi^{(N+1)/2}}{(N+1)\Gamma((N+1)/2)}\right)+\frac{N+1}{2}
     \log\left[
     \frac{4\pi e M^2}{\gm(N+1)}
     \left(b+\frac{N+1}{\max{4/\gm,1/M}}\right)\right].
     \]
\end{remark}

\section{Numerical experiments}\label{Section4}
In this section, we investigate the numerical performance of the SGLD algorithm \eqref{SGLD} for the CVaR minimization problem \eqref{CVaRproblem} from several complementary perspectives. We first assess the accuracy of SGLD on low-dimensional Gaussian portfolios by comparing the resulting portfolio allocations, VaR, and CVaR values with deterministic reference solutions. We then examine how the expected excess risk depends on the number of assets $N$, in line with the dimension dependence suggested by Theorem~\ref{thm2}. Finally, we compare the computational performance of SGLD with several alternative stochastic optimization methods. Together, these experiments illustrate the numerical accuracy, dimension dependence, and computational efficiency of the proposed approach.

\subsection{Simulation results for low-dimensional Gaussian portfolios}

We first consider two low-dimensional Gaussian examples, each consisting of a
portfolio with \(N=3\) assets. We aim to illustrate the
behaviour of SGLD algorithm \eqref{SGLD} in a setting where the reference optimal solution
can be computed accurately by deterministic numerical optimization. This allows
us to compare the portfolio allocation, VaR, and CVaR obtained by SGLD with their
corresponding reference values.

\begin{figure}[h]
	\centering
	\begin{subfigure}{0.48\textwidth}
		\centering
		\includegraphics[width=\textwidth]{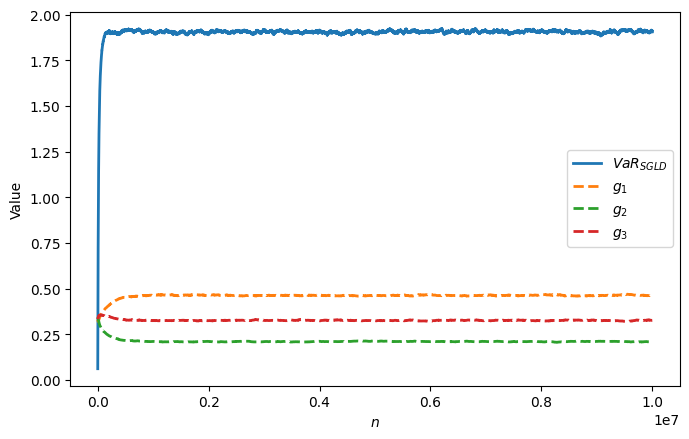}
		\subcaption{\(X_1\sim \mathcal{N}(1,1)\),
					\(X_2\sim \mathcal{N}(0,4)\), \(X_3\sim \mathcal{N}(0.5,2)\)}
	\end{subfigure}
	\hfill
	\begin{subfigure}{0.48\textwidth}
		\centering
		\includegraphics[width=\textwidth]{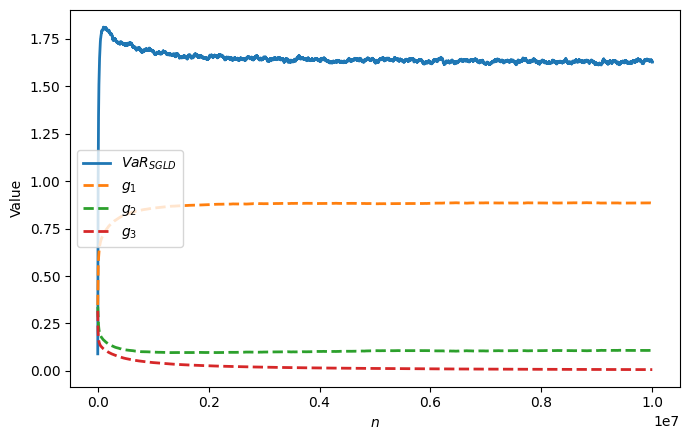}
		\subcaption{\(X_1\sim \mathcal{N}(0,1)\),
                    \(X_2\sim \mathcal{N}(1,4)\), \(X_3\sim \mathcal{N}(3,10^{-4})\)}
	\end{subfigure}
	\caption{Sample paths of the SGLD algorithm for $\text{VaR}_{\text{SGLD}}$, $g_1$, $g_2$, and $g_3$ of the optimization problem \eqref{CVaRproblem} corresponding to the portfolios presented in Table \ref{tab:three_asset_gaussian_results}.}
	\label{fig1}
\end{figure}

In each example, the asset losses $
X_i \sim \mathcal{N}(\mu_i,\sigma_i^2)$, $i=1,2,3$, are assumed to be independent Gaussian random
variables. In Table~\ref{tab:three_asset_gaussian_results}, we compute the reference $95\%$ VaR and CVaR values by applying deterministic optimization to the CVaR objective \eqref{CVaRproblem} and then run the SGLD
algorithm \eqref{SGLD} with $\bar q=0.95$, $\gamma=5\times10^{-5}$,  $\beta=10^{7}$, $\lambda=10^{-5}$, $n=10^7$, and $\Th_0=0\in\R^{N+1}$. To further illustrate the convergence behaviour, Figure~\ref{fig1}
plots sample paths of the SGLD algorithm for the two examples.
\begin{table}[htbp]
	\centering
\caption{\(95\%\) VaR and CVaR for portfolios of three assets \(X_1, X_2, X_3\).}
	\label{tab:three_asset_gaussian_results}	\begin{adjustbox}{max width=\textwidth,center}
		\renewcommand{\arraystretch}{1.5}
		\begin{tabular}{@{}lcccc@{}}
		\toprule 
			Loss distributions 
			& Method 
			& \(g=(g_1,g_2,g_3)\) 
			& VaR 
			& CVaR \\
			\midrule
			
			\multirow{2}{*}{
				\makecell[l]{
					\(X_1\sim \mathcal{N}(1,1)\)\\
					\(X_2\sim \mathcal{N}(0,4)\)\\
					\(X_3\sim \mathcal{N}(0.5,2)\)
			}}
			& Reference
			& \((0.4637,0.2102,0.3261)\)
			& \(1.9056\)
			& \(2.2304\) \\
			
			& SGLD algorithm
			& \((0.4631,0.2102,0.3267)\)
			& \(1.9042\)
			& \(2.2310\) \\
			
			\midrule
			
			\multirow{2}{*}{
				\makecell[l]{
					\(X_1\sim \mathcal{N}(0,1)\)\\
                    \(X_2\sim \mathcal{N}(1,4)\)\\
                    \(X_3\sim \mathcal{N}(3,10^{-4})\)
			}}
			& Reference
			& \((0.8885,0.1110,0.0004)\)
			& \(1.6188\)
			& \(2.0015\) \\
			
			& SGLD algorithm
			& \((0.8850,0.1082,0.0068)\)
			& \(1.6278\)
			& \(2.0099\) \\
			\bottomrule
		\end{tabular}
	\end{adjustbox}
\end{table}

We next consider a moderately higher-dimensional example consisting of $N=15$ independent Gaussian asset losses $X_i\sim\mathcal N(\mu_i,\sigma_i^2),\,i=1,\ldots,15$. Their distributions are specified through the mean and variance vectors given by
$
\mu =
(0,\ 0.3,\ 0.6,\ 1.0,\ 0.2,\ 0.8,\ 1.5,\ 2.0,\ 0.5,\ 1.2,1.8,\ 2.5,\ 0.1,\ 0.9,\ 1.4)$ and $\sigma^2=
(1,\ 1.5,\ 0.8,\ 3.0,\ 4.0,\ 2.0,\ 0.5,\ 1.2,\ 2.5,\ 0.7, 4.5,\ 0.6,\ 3.5,\ 1.0,\ 2.2)$,
respectively. 
We use the same parameter values as in the preceding three-asset experiments. Table~\ref{tab:15_asset_gaussian_results} reports the three assets receiving the largest portfolio weights under each method, together with the corresponding $95\%$ VaR and CVaR values. The results show that the SGLD algorithm identifies the same principal assets as the deterministic reference solution and assigns similar weights to them. Moreover, the VaR and CVaR values produced by SGLD are close to their reference counterparts. This experiment indicates that the agreement observed in the three-asset examples is maintained in a moderately higher-dimensional portfolio.

\begin{table}[h]
	\centering
\caption{\(95\%\) VaR and CVaR for the $15$-asset Gaussian portfolio.}
	\label{tab:15_asset_gaussian_results}	\begin{adjustbox}{max width=\textwidth,center}
		\renewcommand{\arraystretch}{1.5}
		\begin{tabular}{@{}lcccc@{}}
		\toprule 
			Method
            & Largest three weights
            & VaR
            & CVaR\\
			\midrule
			 Reference
			& \((0.2775,0.1422,0.1866)\)
			& \(1.1160\)
			& \(1.3001\) \\
			
			 SGLD algorithm
			& \((0.2768,0.1406,0.1860)\)
			& \(1.1211\)
			& \(1.3038\) \\
			\bottomrule
		\end{tabular}
	\end{adjustbox}
\end{table}

In the next subsection, we further increase the number of assets and investigate the
dimension dependence of the expected excess risk.

\subsection{Dimension dependence under Gaussian loss distributions}
We now investigate how the expected excess risk depends on the number of assets in the portfolio. The experiment is performed for
\begin{equation}\label{dimension}
N\in \{30, 38, 48, 62, 78, 99, 126, 160\}.
\end{equation}
We first generate a collection of $160$ parameter pairs $\mathcal{C}_{160}
:= \big\{(\mu_i,\sigma_i)\big\}_{i=1}^{160}$, where the parameters are generated independently with $\mu_i\sim\mathcal U([0,1])$ and
$\sigma_i\sim\mathcal U([1,2])$ for
$i=1,\ldots,160$, and keep this collection fixed throughout the experiment. For each $N$, the first $N$ parameter pairs in $\mathcal{C}_{160}$ are used to define the multivariate Gaussian loss vector
\[
X=(X_1,\ldots,X_N)^\top\sim\mathcal N(\mu,\Sigma),
\]
where
$\mu=(\mu_1,\ldots,\mu_N)^\top$ and 
$\Sigma=\operatorname{diag}(\sigma_1^2,\ldots,\sigma_N^2)$.

To compute the expected excess risk associated with the CVaR objective \eqref{CVaR}, we use the fact that any linear combination of Gaussian asset losses remains Gaussian. Consequently, both the objective $V$ and its gradient admit explicit deterministic expressions. For each $N$, we solve the resulting first-order optimality conditions to obtain the reference optimum
\[
V^\star=\inf_{\Th\in\mathbb{R}^{N+1}}V(\Th).
\]
Then, we run the SGLD algorithm \eqref{SGLD} with fixed parameters $\bar q=0.95$, $\gamma=5\times10^{-5}$,
$\beta=10^7$, $\lambda=5\times10^{-4}$, $n=2\times10^6$, and $\Th_0=0 \in \R^{N+1}$ to obtain $\Th_n^\lambda$. Then, the estimated expected objective value is obtained by averaging \(V(\Th_n^\lambda)\) over $100$ independent chains. Thus, the expected excess risk is approximated by
\begin{equation}\label{simulateEER}
\mathbb{E}[V(\Th_n^\lambda)]-\inf_{\Th\in\mathbb{R}^{N+1}}V(\Th)
\approx
\frac{1}{100}\sum_{m=1}^{100}V(\Th_{n,m}^{\lambda})-V^\star.
\end{equation}
\begin{figure}[h]
    \centering
    \begin{subfigure}{0.42\textwidth}
      \centering
      \includegraphics[width=\textwidth]{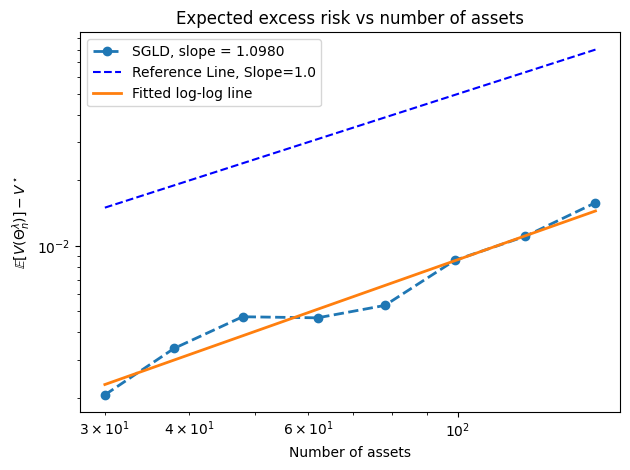}
    \end{subfigure}
       \begin{subfigure}{0.42\textwidth}
      \centering
      \includegraphics[width=\textwidth]{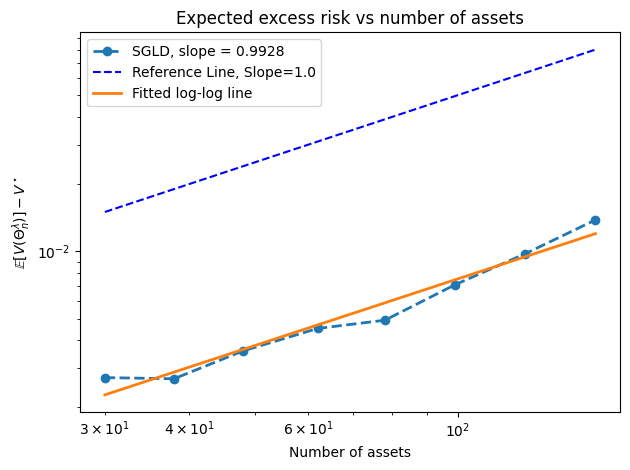}
      \end{subfigure}
    \caption{Expected excess risk versus the number of assets under Gaussian loss distributions.}
    \label{fig2}
\end{figure}
In Figure~\ref{fig2}, we plot the estimated expected excess risk \eqref{simulateEER} against the number of assets $N$ in \eqref{dimension} on a log--log scale, where the fitted slope of the SGLD curve is approximately one, which shows that the expected excess risk grows almost linearly with the number of assets. This numerical observation is consistent with the dimension-dependent upper bound in Theorem~\ref{thm2}.

\subsection{Computational comparison with alternative methods}
In this section, we further compare the computational performance of SGLD with three alternative methods: tSGHMC~\cite{wang2026tamed}, the proximal stochastic subgradient method~\cite{Proximalstochastic}, and PSGLA~\cite{ProximalMCMC}. The purpose of this experiment is to assess not only the number of iterations required to achieve a prescribed level of accuracy, but also the corresponding computational cost.

We consider a portfolio consisting of $N=10$ assets and set the confidence level to $\bar q=0.95$. The regularization parameter, inverse temperature, and stepsize are chosen as $\gamma=5\times10^{-5}$, $\beta=10^7$, and $\lbd = 5 \times 10^{-4}$, respectively. Following the setting of the preceding dimension-dependence experiment, we consider the asset-loss vector
\[
X=(X_1,\ldots,X_{10})^\top\sim\mathcal N(\mu,\Sigma),
\]
where
$\mu=(\mu_1,\ldots,\mu_{10})^\top$ and $\Sigma=\operatorname{diag}(\sigma_1^2,\ldots,\sigma_{10}^2)$ with $\mu_i\sim\mathcal U([0,1])$ and $\sigma_i\sim\mathcal U([1,2])$ generated independently for all $i=1,\ldots,10$.
A fixed random seed is used to generate these parameters, and the same realization of $(\mu,\Sigma)$ is retained across all four algorithms to ensure a consistent and fair comparison. The same initial condition is used for all methods, and the expected objective value at each checkpoint is estimated using $100$ independent trajectories. The reference value $V^*$ is computed by solving the first-order optimality conditions associated with the exact Gaussian CVaR objective.

For all algorithms, we use a target expected excess risk tolerance of
$\ep = 5\times10^{-3}$. The empirical expected excess risk is evaluated every $10^3$ iterations, and an algorithm is regarded as having reached the prescribed accuracy once the estimated excess risk falls below $\ep$. Algorithm-specific parameters are selected as described below. For tSGHMC, we set the friction parameter to $3$, the taming exponent to $r=1$, and the coefficient of the linear term to $m=10^{-3}$. The proximal subproblems arising in the proximal stochastic subgradient method and PSGLA are solved numerically at each iteration. All computations are carried out under the same computational environment so that the reported wall-clock times are directly comparable.

The four methods have substantially different per-iteration computational requirements. SGLD has a particularly simple explicit update, requiring only the evaluation of a stochastic gradient at each iteration. Similarly, tSGHMC remains an explicit scheme, but additionally introduces a momentum variable and requires both position and momentum updates together with the evaluation of the tamed stochastic gradient. In contrast, the proximal stochastic subgradient method and PSGLA require the computation of a proximal step at every iteration, which introduces additional numerical cost. For each method, we record the number of iterations required for the estimated expected excess risk to fall below $5\times10^{-3}$, together with the corresponding wall-clock time. The results are summarized below.

\begin{table}[htbp]
    \centering
    \caption{Computational comparison of SGLD, tSGHMC, the proximal stochastic subgradient method, and PSGLA for a target expected excess risk tolerance of $5\times10^{-3}$.}
    \label{tab:computational_comparison}
    \begin{tabular}{lccc}
        \hline
        Method
        & Iterations to tolerance
        & Final excess risk
        & Wall-clock time (s) \\
        \hline
        SGLD
        & $41,000$
        & $4.8607\times 10^{-3}$
        & $5$ \\
        tSGHMC~\cite{wang2026tamed}
        & $143,000$
        & $4.7941\times 10^{-3}$
        & $22$ \\
        Proximal stochastic subgradient~\cite{Proximalstochastic}
        & $43,000$
        & $4.2746\times 10^{-3}$
        & $1,190$ \\
        PSGLA~\cite{ProximalMCMC}
        & $29,000$
        & $4.3068\times 10^{-3}$
        & $2,748$ \\
        \hline
    \end{tabular}
\end{table}
Table~\ref{tab:computational_comparison} shows that all four methods are able to reach the prescribed expected excess risk tolerance of $5\times10^{-3}$, but their computational requirements differ substantially. SGLD reaches the target after $41,000$ iterations in approximately $5$ seconds. The proximal stochastic subgradient method requires a comparable number of iterations, namely $43,000$, but its computational cost is substantially higher, with a wall-clock time of approximately $1,190$ seconds. PSGLA reaches the prescribed tolerance in the smallest number of iterations, namely $29,000$, however, it requires approximately $2,748$ seconds and is therefore the most computationally expensive method in this comparison. In contrast, tSGHMC requires the largest number of iterations, $143,000$, but remains relatively inexpensive computationally, reaching the target in approximately $22$ seconds.

These results demonstrate that the two proximal methods require more computational time than SGLD, despite reaching the prescribed expected excess risk tolerance with a comparable or even smaller number of iterations. Although tSGHMC requires more iterations than SGLD, its explicit structure keeps the computational cost relatively moderate. Among the four methods considered, SGLD achieves the prescribed accuracy with the lowest wall-clock time, suggesting a favorable balance between empirical accuracy and computational efficiency for the CVaR portfolio optimization problem considered here.

\section{General case}\label{Section5}
In Section \ref{CVaR}, we applied the SGLD algorithm to the specific CVaR optimization problem \eqref{CVaRproblem}. In this section, we consider a general non-convex stochastic optimization problem of the form
\begin{equation}\label{optimization}
\text{minimize} \quad \R^d \ni \Th \mapsto U(\Th) := \E[f(\Th,X)],
\end{equation}
where $X$ is an $\R^m$-valued random variable with probability law $\cL(X)$ and $f : \R^d \times \R^m \to \R$ is a measurable function satisfying $\E[|f(\Th,X)|] < \infty$ for all
\( \Th \in \mathbb{R}^d \). Let $h:=\nabla U$ denote the gradient of $U$. Moreover, for a fixed inverse temperature parameter $\beta>0$, define
\begin{equation}
	\pi'_\beta(A) :=
	\frac{\int_A e^{-\beta U(\Th)} \, d\Th}
	{\int_{\mathbb{R}^d} e^{-\beta U(\Th)} \, d\Th},
	\qquad
	A \in \mathcal{B}(\mathbb{R}^d),
\end{equation}
with $\int_{\mathbb{R}^d} e^{-\beta U(\Th)} \, d\Th < \infty$.
To solve \eqref{optimization}, we consider the SGLD algorithm given by
\begin{equation}\label{SGLD2}
	\Th_0^{\text{SGLD}} := \Th_0',
	\qquad
	\Th_{n+1}^{\text{SGLD}} :=
	\Th_n^{\text{SGLD}}
	- \lbd H(\Th_n^{\text{SGLD}}, X_{n+1})
	+ \sqrt{2\lbd \beta^{-1}} \, \xi_{n+1},\qquad n\in\N_0,
\end{equation}
where $\Th_0'$ is an $\R^d$-valued random variable, $\lbd >0$ is the stepsize, $\beta >0$,  $(X_n)_{n \in \N_0}$ is an
$\R^m$-valued i.i.d. process with common law $\cL(X)$, $H : \R^d \times \R^m \to \R^d$ is a measurable function, and $(\xi_n)_{n \in \N}$ is an independent sequence of standard $d$-dimensional Gaussian random variables.

Then, we impose the following assumptions.

\begin{assumption}\label{assumption1}
The initial value $\Th_0'$ satisfies $\E|\Th_0'|^2 < \infty$. The i.i.d. process $(X_n)_{n\in\N_0}$ satisfies $\E|X_0|^4 <\infty$. Moreover, $h(\Th)=\E[H(\Th,X_0)]$, for all $\Th \in \R^{d}$.
\end{assumption}

\begin{assumption}\label{assumption2}
There exist measurable functions $\eta_1,\eta_2 : \R^m \rightarrow (0,\infty)$ satisfying $\E[\eta_1(X_0)^2]<\infty$, $\E[\eta_2(X_0)^2]<\infty$, and constants $m_H,L_H > 0$ such that, for all $\Th\in\R^{d}$ and $x\in\R^m$,
\[
|H(\Th,x)|  \leq m_H\eta_1(x)+L_H\eta_2(x)|\Th|.
\]
\end{assumption}

\begin{assumption}\label{assumption3}
There exist a measurable symmetric matrix-valued function
$A_1:\mathbb R^m\to \mathbb R^{d\times d}$ and a constant $r_H>0$ such that the smallest eigenvalue $\upsilon_{\min}\big(\mathbb E[A_1(X_0)]\big):= \gm_H \in (0,\infty)$, and, for all $\Th,\Th'\in\mathbb R^d$ and $x\in\mathbb R^m$,
\[
\left\langle H(\Th,x)-H(\Th',x),\Th-\Th'\right\rangle
\geq
\left\langle \Th-\Th',
A_1(x)(\Th-\Th')\right\rangle,
\qquad
\textit{if }|\Th-\Th'|\geq r_H.
\]
\end{assumption}

\begin{assumption}\label{assumption4}
There exists a measurable positive semidefinite symmetric matrix-valued function
$A_2:\mathbb R^m\to \mathbb R^{d\times d}$ such that
$\|\mathbb E[A_2(X_0)]\|_{\mathrm{op}}:=k_H \in [0,\infty)$,
and, for all $\Th,\Th'\in\mathbb R^d$ and $x\in\mathbb R^m$,
\[
\left\langle H(\Th,x)-H(\Th',x),\Th-\Th'\right\rangle
\geq
-\left\langle \Th-\Th',
A_2(x)(\Th-\Th')\right\rangle.
\]
\end{assumption}
\begin{remark}
The above assumptions are also satisfied by the SCAD-penalized linear least-squares problem~\cite{SCAD}:
\[
\min_{\Th\in\R^d} U(\Th)=
\left\{
\E\big[|Z-\langle Y,\Th\rangle|^2\big]
+
\sum_{j=1}^d p_{\chi,a}(|\Th_j|)
\right\},
\]
where $\Th\in\R^d$ is the parameter, $Y\in\R^d$ is the covariate vector, $Z\in\R$ is the response variable, $\chi>0$, $a>2$, and $p_{\chi,a}$ denotes the SCAD penalty. Let
$
S(\Th)=\big(s(\Th_1),\ldots,s(\Th_d)\big)^\top,
$
where
\[
s(t)=
\begin{cases}
\chi\operatorname{sgn}(t),
& 0<|t|\leq\chi,\\
\dfrac{a\chi-|t|}{a-1}\operatorname{sgn}(t),
& \chi<|t|\leq a\chi,\\
0,
& |t|>a\chi,
\end{cases}
\qquad s(0)=0,
\]
is a measurable selection of the SCAD subgradient. A stochastic generalized gradient of $U$ is then given by
\[
H(\Th,x)=
2y(\langle y,\Th\rangle-z)+S(\Th),
\qquad
x=(y,z)\in\R^d\times\R.
\]
Suppose that $\E|X_0|^4<\infty$ and $\mu_Y:=
\upsilon_{\min}\big(\E[Y_0Y_0^\top]\big)>0$.
Since $|S(\Th)|\leq \chi\sqrt d$, we have
\[
|H(\Th,x)|
\leq 2|y||\langle y, \Th \rangle|+2|y||z|+|S(\Th) | \leq
2|y|^2|\Th|+2|y||z|+\chi\sqrt d,
\]
and hence Assumption~\ref{assumption2} holds by taking $\eta_1(x)=1+|y||z|$, $\eta_2(x)=1+|y|^2$,
together with $m_H = \max\{2,\chi\sqrt{d}\}$ and $L_H=2$. To verify Assumption~\ref{assumption3}, we have
\[
\langle H(\Th,x)-H(\Th',x),\Th-\Th'\rangle
=
2\langle y,\Th-\Th'\rangle^2
+
\langle S(\Th)-S(\Th'),\Th-\Th'\rangle
\geq
2\langle y,\Th-\Th'\rangle^2-
2\chi\sqrt d|\Th-\Th'|.
\]
Therefore, when $|\Th-\Th'|\geq r_H$,
\[
\langle H(\Th,x)-H(\Th',x),\Th-\Th'\rangle
\geq
\left\langle
\Th-\Th',
A_1(x)(\Th-\Th')
\right\rangle,
\]
where
$
A_1(x) =
2yy^\top-
2\chi\sqrt d /r_H I_d$, implying $
\nu_{\min}\big(\E[A_1(X_0)]\big) =
2\mu_Y-2\chi\sqrt d/r_H$.
Thus, Assumption~\ref{assumption3} holds for any
$r_H>\chi\sqrt d/\mu_Y$ and $\gm_H=2\mu_Y-2\chi\sqrt d/r_H
>0.$
Finally, by the definition of $s(t)$, its slope is $0$ on $0<|t|<\chi$, $-1/(a-1)$ on $\chi<|t|<a\chi$, and $0$ on $|t|>a\chi$. Furthermore, $s$ is continuous at $\pm \chi$ and $\pm a\chi$, while its discontinuity at $0$ is an upward jump. Hence, for any $u,v\in\R$,
\[
\frac{s(u)-s(v)}{u-v}
\geq
-\frac{1}{a-1},
\qquad u\neq v.
\]
Multiplying both sides by $|u-v|^2$ gives
\[
\big(s(u)-s(v)\big)(u-v)
\geq
-\frac{1}{a-1}|u-v|^2,
\qquad u,v\in\R.
\]
Consequently,
\[
\langle S(\Th)-S(\Th'),\Th-\Th'\rangle
=
\sum_{j=1}^d
\big(s(\Th_j)-s(\Th_j')\big)
(\Th_j-\Th_j')
\geq
-\frac{1}{a-1}|\Th-\Th'|^2.
\]
Moreover, since 
\[
\big\langle 2y(\langle y,\Th \rangle-z) -2y(\langle y,\Th' \rangle-z),\Th-\Th'\big\rangle =2 \big\langle y\langle y,\Th-\Th' \rangle,\Th-\Th'\big\rangle = 2\langle y,\Th-\Th' \rangle^2 \geq 0,
\]
we obtain that
\[
\langle H(\Th,x)-H(\Th',x),\Th-\Th'\rangle
\geq
-\frac{1}{a-1}|\Th-\Th'|^2.
\]
Therefore Assumption~\ref{assumption4} holds by taking
$
A_2(x)=1/(a-1) I_d$ and $k_H= \|\E[A_2(X_0)]\|_{\mathrm{op}}=1/(a-1).
$
\end{remark}

Under these assumptions, we extend the Wasserstein convergence result for the SGLD algorithm to the general stochastic optimization framework, and then derive the corresponding expected excess risk bound for the objective function \(U\).

\begin{theorem}\label{thm3}
 Let Assumptions \ref{assumption1}--\ref{assumption4} hold and set $\tl{\lbd}_0=\min\{\gm_H/\big(2L_H^2\E[\eta_2(X_0)^2]\big)$, $1$, $1/\gm_H\}$. Then, for any $\beta>0$, there exist constants $\tl{C}_5$, $\tl{C}_6$, and $\tl{T}_0>0$ such that, for any $\lbd \in(0,\tl{\lbd}_0)$ and $n \in \N_0$,
\[
W_2(\cL(\Th_{n}^{\mathrm{SGLD}}),\pi_\beta') \leq  2\tl{C}_5 \lbd^{1/4} +e\tl{C}_6 e^{-n\lbd/\tl{T}_0}.
\]
where $\tl{T}_0$ are independent of the dimension $d$, $\tl{C}_5 = \mathcal{O}(\sqrt{d})$, and $\tl{C}_6 = \mathcal{O}(\sqrt{d})$.
\end{theorem}
\begin{proof}
The proof follows the same argument as the proof of Theorem \ref{thm1} for the CVaR problem, with the objective function $V$, the dimension $N+1$, and the constants $m,L,\gm,r,k$ replaced respectively by $U$, $d$, and the corresponding deterministic constants together with the moment bounds of $\eta_1(X_0)$ and $\eta_2(X_0)$, as induced by Assumptions~\ref{assumption1}--\ref{assumption4}.
\end{proof}

\begin{theorem}\label{thm4}
 Let Assumptions \ref{assumption1}--\ref{assumption4} hold and set $\tl{\lbd}_0=\min\{\gm_H/\big(2L_H^2\E[\eta_2(X_0)^2]\big)$, $1$, $1/\gm_H\}$. Then, for any $\beta>\max\{4/\gm_H, 1/\tl{M}\}$ with $\tl{M} = 3L_H\E[\eta_2(X_0)]/2+m_H\E[\eta_1(X_0)]+L_H\E[\eta_2(X_0)]\sqrt{2b_H/\gm_H}$ and $b_H:= \max\{\lf(m_H\E[\eta_1(X_0)]\rt)^2/(2\gm_H), m_Hr_H+(L_H\E[\eta_2(X_0)]+\gm_H/2)r_H^2\}$, there exists constants $\tl{C}_V$, $\tl{C}_5$, $\tl{C}_6$, and $\tl{T}_0 >0$ such that, for any $\lbd \in(0,\tl{\lbd}_0)$ and $n \in \N_0$,
\begin{align*}
	\E[U(\Th_n^{\mathrm{SGLD}})]-\inf_{\Th \in \R^{d}}U(\Th)&\leq \tl{C}_{V}\lf(2\tl{C}_5 \lbd^{1/4} +e\tl{C}_6 e^{-n\lbd/\tl{T}_0}  \rt) +\frac{2}{\beta}-\frac{1}{\beta}\log\lf(\frac{2\pi^{d/2}}{d \Gamma(d/2)}\rt)\\
	&\quad +	\frac{d}{2\beta}
	\log\left(
	\frac{4\pi e(b_H+d/\beta)\beta^2\tl{M}^2}{\gm_H d}
	\right),
\end{align*}
  where $\tl{T}_0$ are independent of $d$, $\tl{C}_V= \mathcal{O}(\sqrt{d})$, $\tl{C}_5= \mathcal{O}(\sqrt{d})$, and $\tl{C}_6= \mathcal{O}(\sqrt{d})$.
\end{theorem}
\begin{proof}
	The proof follows the same argument as the proof of Theorem \ref{thm2}. More precisely, following the proofs of Lemmas \ref{CV1} and \ref{CV2}, one obtains
	\[
	\E[U(\Th^{\text{SGLD}}_n)]-\E[U(\Th'_\infty)]  \leq \tl{C}_{V}W_2(\cL(\Th^{\text{SGLD}}_n), \pi'_\beta),	
	\]
	where $\Th'_\infty \sim \pi'_\beta$, and
	\[
	\E[U(\Th'_\infty)]-\inf_{\Th \in \R^{d}} U(\Th)  \leq \frac{d}{2\beta}
	\log\left(
	\frac{4\pi e(b_H+d/\beta)\beta^2\tl{M}^2}{\gm_H d}
	\right)
	+\frac{2}{\beta}
	-
	\frac{1}{\beta}\log\lf(\frac{2\pi^{d/2}}{d \Gamma(d/2)}\rt).
	\]
	Combining these two estimates with Theorem \ref{thm3} gives the desired bound.
\end{proof}

\newpage
\appendix

\makeatletter
\renewcommand{\thelemma}{\arabic{lemma}}
\renewcommand{\theproposition}{\arabic{proposition}}
\allowdisplaybreaks

\section{Auxiliary lemmas}\label{Auxiliarylemmas}
In this section, we establish several auxiliary lemmas that will be used in the proofs of the main theorems. Recall the definition of $H$ in \eqref{Hetawj}. For any $\Th\in\R^{N+1}$ and $x \in \R^N$, define $I(\Th,x):=\mathbbm{1}_{\{G(w,x)\geq\eta\}}$ and let $R(\Th,x):=(R_\eta(\Th,x),R_{w_1}(\Th,x),\ldots,R_{w_N}(\Th,x))^\top$ with
\begin{equation}\label{R(theta,x)}
	R_\eta(\Th,x):=-I(\Th,x), \hspace{5mm} R_{w_j}(\Th,x):=\hat{g}_{w_j}(w,x) I(\Th,x), \quad j=1,\ldots,N.
\end{equation}
Then, $H(\Th,x)$ can be written as
\begin{equation}\label{H(theta,x)}
	H(\Th,x)= e_1+\frac{1}{1-\bar{q}}R(\Th,x)+2\gm \Th,
\end{equation}
where $e_1=(1,0,0,\ldots,0)^\top \in \R^{N+1}$.

Moreover, for any $\Th\in\R^{N+1}$ and $x \in \R^N$, define
\[
F(\Th,x):=G(w,x)-\eta.
\]
Note that $F$ is twice differentiable everywhere in $\Th$ and its gradient $\nabla_\Th F$ is given by
\[
\nabla_\Th F(\Th,x)=\lf(-1,\hat{g}_{w_1}(w,x),\ldots,\hat{g}_{w_N}(w,x)\rt)^\top.
\]
By definition \eqref{R(theta,x)}, we have $R(\Th,x) =I(\Th,x) \nabla_\Th F(\Th,x)$.

Furthermore, for any $\Th\in\R^{N+1}$ and $x \in \R^N$, we define
\begin{equation}\label{Phi}
\Phi(\Th,x):= \max\{F(\Th,x)+\frac{\mathcal{M}(x)}{2}|\Th|^2,\frac{\mathcal{M}(x)}{2}|\Th|^2\} \tag{$\Phi$}
\end{equation}
where $\mathcal{M}(x) := 2 \max_{i=1,\ldots,N} |x^i|$. Note that, for fixed $x \in \R^N$, \(\Phi(\cdot,x)\) may not be differentiable at points where \(F(\Th,x)=0\), we therefore define the subdifferential $\partial_\Th \Phi(\Th,x)$ of $\Phi$ at $\Th$ by 
\[
\partial_\Th \Phi(\Th,x) :=\lf\{p\in\R^{N+1}: \Phi(u,x)\geq \Phi(\Th,x)+\langle p,u-\Th\rangle \;\;\forall\, u\in \R^{N+1}\rt\}.
\]

\begin{remark}
    Assumption~\ref{A1}, in particular $\E|X_0|^2<\infty$, implies that $\E[\mathcal{M}(X_0)^2]<\infty$.
\end{remark}

With the above definitions, we present below in Lemmas \ref{A2}--\ref{A4} several properties of $H$ that are key in establishing convergence results for SGLD \eqref{SGLD}.

\begin{lemma}\label{A2}
	Suppose Assumption~\ref{A1} holds. Then, for all $\Th\in\R^{N+1}$ and $x\in \R^{N}$,
	\[
	|H(\Th,x)|^2 \leq 2\left(1+\frac{1}{1-\bar q}\right)^2+\frac{2\mathcal{M}(x)^2}{(1-\bar q)^2}+8\gm^2|\Th|^2.
	\]
 Moreover, for all $\Th\in\R^{N+1}$,
	\[
	|h(\Th)|^2 \leq m+L|\Th|^2,
	\]
    where
	$m=2\left(1+\frac{1}{1-\bar q}\right)^2+\frac{2\E[\mathcal{M}(X_0)^2]}{(1-\bar q)^2}$ and $L=8\gm^2$.
\end{lemma}
\begin{proof}
	Fix $x \in \R^N$. By the definition of $g_i(w)$ in \eqref{g_i(w)}, we have, for all $w \in \R^{N}$, that
	\[
	\frac{\partial g_i(w)}{\partial w_j}=g_i(w)\bigl(\mathbbm{1}_{\{i=j\}}-g_j(w)\bigr).
	\]
	It follows that
	\begin{equation}\label{gwj}
	\hat g_{w_j}(w,x)=\frac{\partial G(w,x)}{\partial w_j}
	=g_j(w)\left(x^j-\sum_{i=1}^Ng_i(w)x^i\right).\tag{$\star$}
	\end{equation}
	Hence, by \eqref{Hetawj}, we have, for all $\Th \in \R^{N+1}$, that
	\[
	|H_\eta(\Th,x)|^2
	\leq 2\lf(1+\frac{1}{1-\bar{q}}\rt)^2+8\gm^2|\eta|^2,
	\]
	and
	\begin{align*}
		|H_{w_j}(\Th,x)|^2& \leq \frac{2}{(1-\bar{q})^2} |\hat g_{w_j}(w,x)|^2+8\gm^2|w_j|^2\\
		&=\frac{2}{(1-\bar{q})^2}g_j(w)^2\left|x^j-\sum_{i=1}^Ng_i(w)x^i\right|^2+8\gm^2|w_j|^2 \\
		&\leq \frac{2}{(1-\bar{q})^2}\bigg(2g_j(w)^2|x^j|^2
		+2g_j(w)^2\left|\sum_{i=1}^Ng_i(w)x^i\right|^2\bigg) +8\gm^2|w_j|^2\\
		&\leq \frac{2}{(1-\bar{q})^2}\bigg(2g_j(w)^2|x^j|^2
		+2g_j(w)^2\sum_{i=1}^Ng_i(w)|x^i|^2\bigg) +8\gm^2|w_j|^2,
	\end{align*}
	where the last inequality follows from Jensen's inequality. Since $\sum_{i=1}^Ng_i(w)^2 \leq \sum_{i=1}^Ng_i(w)=1$, we obtain
	\begin{align*}
		|H(\Th,x)|^2 &= |H_\eta(\Th,x)|^2+\sum_{j=1}^N|H_{w_j}(\Th,x)|^2 \nonumber\\
		&\leq 2\lf(1+\frac{1}{1-\bar{q}}\rt)^2
		+\frac{2}{(1-\bar{q})^2}\sum_{j=1}^N\bigg(2g_j(w)^2|x^j|^2
		+2g_j(w)^2 \max_{i=1,\ldots,N}|x^i|^2\bigg)+8\gm^2|\Th|^2 \nonumber\\
		&\leq 2\lf(1+\frac{1}{1-\bar{q}}\rt)^2 +\frac{8}{(1-\bar{q})^2}  \max_{i=1,\ldots,N}|x^i|^2+8\gm^2|\Th|^2\nonumber\\
		&=2\lf(1+\frac{1}{1-\bar{q}}\rt)^2 +\frac{2\mathcal{M}(x)^2}{(1-\bar{q})^2}  +8\gm^2|\Th|^2,
	\end{align*}
	where $\mathcal{M}(x) := 2\max_{i=1,\ldots,N}|x^i|$. Taking expectation with respect to $X_0$ gives
	\[
	\E|H(\Th,X_0)|^2 \leq m  +L|\Th|^2,
	\]
	where $m=2(1+\frac{1}{1-\bar{q}})^2 +\frac{2\E[\mathcal{M}(X_0)^2]}{(1-\bar{q})^2}$ and $L =8\gm^2 $. Finally, by Assumption \ref{A1} and Jensen's inequality,
	\[
	|h(\Th)|^2 =\lf| \E[H(\Th,X_0)]\rt|^2 \leq \E|H(\Th,X_0)|^2 \leq m  +L|\Th|^2,
	\]
which completes the proof.
\end{proof}

\begin{lemma}\label{A3}
	Suppose Assumption~\ref{A1} holds. Then, for all $\Th,\Th'\in\R^{N+1}$,
	\[
	\left\langle h(\Th)-h(\Th'),\Th-\Th'\right\rangle
	\geq \gm|\Th-\Th'|^2,
	\qquad \text{if} \;\; |\Th-\Th'|\geq r,
	\]
	where $r =  2(1+\E[\mathcal{M}(X_0)])/(\gm(1-\bar{q}))$.
\end{lemma}

\begin{proof} Fix $x \in \R^N$. Recall the definitions of $R$ in \eqref{R(theta,x)} and \eqref{gwj}. Since $\sum_{i=1}^Ng_i(w)=1$, we have, for all $\Th \in \R^{N+1}$, that
	\begin{align*}
		|R(\Th,x)|&\leq|R_\eta(\Th,x)|+ \sum_{j=1}^N|R_{w_j}(\Th,x)|\\
		&\leq 1+\sum_{j=1}^Ng_j(w)\big(|x^j|+\max_{i=1,\ldots,N}|x^i|\big)\\
		&\leq 1+\mathcal{M}(x).
	\end{align*}
	Combining this estimate with \eqref{H(theta,x)} yields, for all
	$\Th,\Th'\in\R^{N+1}$, that
	\begin{align*}
		\left\langle H(\Th,x)-H(\Th',x),\Th - \Th'\right\rangle &= \frac{1}{1-\bar{q}}\left\langle R(\Th,x)-R(\Th',x),\Th - \Th'\right\rangle +2\gm|\Th-\Th'|^2\\
		&\geq -\frac{1}{1-\bar{q}}\left|R(\Th,x)-R(\Th',x)\right||\Th-\Th'| +2\gm|\Th-\Th'|^2\\
		&\geq -\frac{1}{1-\bar{q}}(|R(\Th,x)|+|R(\Th',x)|)|\Th-\Th'|+2\gm|\Th-\Th'|^2 \\
		&\geq -\frac{2(1+\mathcal{M}(x))}{1-\bar{q}}|\Th-\Th'|+2\gm|\Th-\Th'|^2,
	\end{align*}
which further implies that
	\begin{align*}
    \left\langle h(\Th)-h(\Th'),\Th - \Th'\right\rangle &=
	\left\langle \E[H(\Th,X_0)-H(\Th',X_0)],\Th - \Th'\right\rangle \\
    &\geq -\frac{2(1+\E[\mathcal{M}(X_0)])}{1-\bar{q}}|\Th-\Th'|+2\gm|\Th-\Th'|^2.
	\end{align*}
	If $|\Th-\Th'| \geq r:= 2(1+\E[\mathcal{M}(X_0)])/\big(\gm(1-\bar{q})\big)$, then
	\begin{align*}
		-\frac{2(1+\E[\mathcal{M}(X_0)])}{1-\bar{q}}|\Th-\Th'|+2\gm|\Th-\Th'|^2 =\lf[-\frac{2(1+\E[\mathcal{M}(X_0)])}{(1-\bar{q})|\Th-\Th'|}+2\gm\rt]|\Th-\Th'|^2 \geq \gm|\Th-\Th'|^2.
	\end{align*}
	Therefore, we have
	\[
	\left\langle h(\Th)-h(\Th'),\Th - \Th'\right\rangle
	\geq \gm|\Th - \Th'|^2,
	\qquad \text{if} \;|\Th - \Th'|\geq r,
	\]
which completes the proof.
\end{proof}

\begin{lemma}\label{dissipativity}
	Suppose Assumption~\ref{A1} holds. Then, for all $\Th\in\R^{N+1}$,
	\[
	\langle h(\Th),\Th\rangle
	\geq \frac{\gm}{2}|\Th|^2-b,
	\]
	where $
	b:=\max\left\{
	m/(2\gm),\;
	\sqrt{m}r+\left(\sqrt{L}+\gm/2\right)r^2
	\right\}$.
\end{lemma}
\begin{proof}
    First suppose that $|\Th| \geq r$. By Young's inequality, Lemmas \ref{A2} and \ref{A3}, we have
	\begin{align*}
	\left\langle h(\Th),\Th \right\rangle =\left\langle \E[H(\Th,X_0)],\Th \right\rangle &=
	\left\langle \E[H(\Th,X_0)-H(0,X_0)],\Th \right\rangle +\left\langle \E[H(0,X_0)],\Th \right\rangle \\
    &\geq \gm|\Th |^2 -\frac{\gm|\Th|^2}{2}-\frac{|\E[H(0,X_0)]|^2}{2\gm}\\
    &\geq \frac{\gm|\Th|^2}{2}-\frac{m}{2\gm}.
\end{align*}
Next suppose that $|\Th| < r$. By Lemma~\ref{A2}, we have
\begin{align*}
    \left\langle \E[H(\Th,X_0)],\Th \right\rangle \geq -|\Th||\E[H(\Th,X_0)]|&\geq -|\Th|(\sqrt{m}+\sqrt{L}|\Th|) \\
    &\geq -\sqrt{m}r-\sqrt{L}r^2+\frac{\gm}{2}r^2-\frac{\gm}{2}r^2\\
    &\geq \frac{\gm}{2}|\Th|^2 -\lf(\sqrt{m}r+\lf(\sqrt{L}+\frac{\gm}{2}\rt)r^2\rt).
\end{align*}
Combining the two cases, we obtain, for all $\Th \in \R^{N+1}$,
\[
\langle h(\Th),\Th\rangle
\geq \frac{\gm}{2}|\Th|^2-b,
\]
where $b= \max\{m/(2\gm), \sqrt{m}r+(\sqrt{L}+\gm/2)r^2\}$.
\end{proof}

\begin{lemma}\label{A4}
Suppose Assumption~\ref{A1} holds. Then, for all $\Th,\Th'\in\R^{N+1}$,
	$$
	\langle h(\Th)-h(\Th'),\Th-\Th'\rangle\geq -k|\Th-\Th'|^2,
	$$
	where $k= \lf(\E[\mathcal{M}(X_0)]/(1-\bar{q})-2\gm\rt)_+$.
\end{lemma}
\begin{proof}
	Fix $x\in \R^N$. We claim the following two statements:\\
	\textbf{Claim 1.} $\Th \mapsto F(\Th,x)$ is a $\mathcal{M}(x)$-semi-convex function, i.e., the function $\Th \mapsto F(\Th,x)+\frac{\mathcal{M}(x)}{2}|\Th|^2$ is convex. \\
	\textbf{Claim 2.} $R(\Th,x)+\mathcal{M}(x)\Th \in \partial_\Th \Phi(\Th,x)$ for all $\Th \in \R^{N+1}$.
    
	Recall $\Phi$ defined in \eqref{Phi}. We note that, by Claim 1, $\Phi$ is convex in $\Th$, and, by Claim 2, $R(\Th,x)+\mathcal{M}(x)\Th$ is a subdifferential of $\Phi$. Then, by the fact that any subdifferential of a convex function is monotone, we obtain, for all $\Th,\Th' \in \R^{N+1}$ and $x\in\R^N$, that
	\[
	\langle R(\Th,x)+\mathcal{M}(x)\Th - (R(\Th',x)+\mathcal{M}(x)\Th'),\Th-\Th'\rangle \geq0,
	\]
	which implies that
	\[
	\langle R(\Th,x) - R(\Th',x),\Th-\Th'\rangle \geq -\mathcal{M}(x)|\Th-\Th'|^2.
	\]
	Recall the definition of $H$ in \eqref{H(theta,x)}, we have
	\begin{align}\label{H(v,x)-H(v',x)}
		\langle H(\Th,x)-H(\Th',x),\Th-\Th'\rangle &= \frac{1}{1-\bar{q}}\langle R(\Th,x)-R(\Th',x),\Th-\Th'\rangle+2\gm|\Th-\Th'|^2\nonumber\\
		&\geq \lf(2\gm-\frac{\mathcal{M}(x)}{1-\bar{q}}\rt)|\Th-\Th'|^2,
	\end{align}
which implies
	\begin{align*}
		\langle h(\Th)-h(\Th'),\Th-\Th'\rangle = \langle \E[H(\Th,X_0)-H(\Th',X_0)],\Th-\Th'\rangle &\geq \lf(2\gm-\frac{\E[\mathcal{M}(X_0)]}{1-\bar{q}}\rt)|\Th-\Th'|^2\\
		&\geq -k|\Th-\Th'|^2,
	\end{align*}
	where $k = \lf(\E[\mathcal{M}(X_0)]/(1-\bar{q})-2\gm\rt)_+$.
	
	\vspace{2mm}
	Then, it remains to prove the claims.
	
	\noindent\textit{Proof of Claim 1.} Recall that $F(\Th,x)=G(w,x)-\eta$ which is twice continuously differentiable. To prove $\Th \mapsto F(\Th,x)$ is $\mathcal{M}(x)$-semi-convex, it suffices to show \[
    z^\top\nabla^2_w G(w,x)z + \mathcal{M}(x)|z|^2 \geq 0\]
    for all $w,z \in \R^N$. Fix $w,z \in \R^N$. Define $\bar{z} := \sum_{i=1}^N g_i(w)z_i$ and $f:\R \rightarrow \R$ by $f(t)=G(w+tz,x)$, then
	\begin{equation}\label{f''(0)}
		f''(0) = z^\top\nabla^2_w G(w,x)z.
	\end{equation}
	We further define 
	\[
	\tilde{g}_i(t):=g_i(w+tz)=\frac{e^{w_i+tz_i}}{\sum_{l=1}^N e^{w_l+tz_l}}\in(0,1), \;i=1,\ldots,N.
	\]
	Then, we have, for all $t\in \R$, that
	\[
	\tilde{g}'_i(t)=\tilde{g}_i(t)(z_i-\sum_{j=1}^N \tilde{g}_j(t)z_j) \qquad \text{and} \qquad f(t)=\sum_{i=1}^N g_i(w+tz)x^i=\sum_{i=1}^N \tilde{g}_i(t)x^i.
	\]
	Let $\tilde{f}(t):= \sum_{i=1}^N \tilde{g}_i(t)z_i$, we get
	\[
	f'(t)=\sum_{i=1}^N \tilde{g}'_i(t)x^i=\sum_{i=1}^N x^i\tilde{g}_i(t)(z_i-\sum_{j=1}^N \tilde{g}_j(t)z_j)=\sum_{i=1}^N x^i\tilde{g}_i(t)(z_i-\tilde{f}(t)).
	\]
	Moreover, we obtain
	\[
	f''(t) = \sum_{i=1}^N \tilde{g}''_i(t)x^i =\sum_{i=1}^N x^i\lf(\tilde{g}_i(t)(z_i-\tilde{f}(t))^2-\tilde{g}_i(t) \tilde{f}'(t)\rt),
	\]
	where
	\begin{align*}
		\tilde{f}'(t) = \sum_{i=1}^N \tilde{g}'_i(t)z_i&=\sum_{i=1}^N \tilde{g}_i(t)(z_i-\tilde{f}(t))( z_i-\tilde{f}(t)+\tilde{f}(t))\\
		&=\sum_{i=1}^N \tilde{g}_i(t)(z_i-\tilde{f}(t))^2+\tilde{f}(t)\sum_{j=1}^N \tilde{g}_j(t)(z_j-\tilde{f}(t))\\
		&=\sum_{i=1}^N \tilde{g}_i(t)(z_i-\tilde{f}(t))^2+\tilde{f}(t)\bigg(\tilde{f}(t)-\sum_{j=1}^N \tilde{g}_j(t)\tilde{f}(t)\bigg)\\
		&=\sum_{i=1}^N \tilde{g}_i(t)(z_i-\tilde{f}(t))^2,
	\end{align*}
	where the last equality holds due to $\sum_{j=1}^N\tilde{g}_j(t)=1.$
	Then, we have
	\begin{align*}
		f''(t) &=\sum_{i=1}^N x^i\lf(\tilde{g}_i(t)(z_i-\tilde{f}(t))^2-\tilde{g}_i(t)\sum_{j=1}^N \tilde{g}_j(t)(z_j-\tilde{f}(t))^2\rt)\\
		& =\sum_{i=1}^N \tilde{g}_i(t) \lf(x^i - \sum_{j=1}^N \tilde{g}_j(t)x^j\rt)(z_i-\tilde{f}(t))^2,
	\end{align*}
	which implies that
	\[
	f''(0) =\sum_{i=1}^N \tilde{g}_i(0) \lf(x^i - \sum_{j=1}^N \tilde{g}_j(0)x^j\rt)(z_i-\tilde{f}(0))^2=\sum_{i=1}^N g_i(w) \lf(x^i - \sum_{j=1}^N g_j(w)x^j\rt)(z_i-\bar{z})^2.
	\]
	By \eqref{f''(0)}, we get
	\begin{align*}
		|z^\top\nabla^2_w G(w,x)z|&\leq \max_{i=1,\ldots,N}\bigg|x^i-  \sum_{j=1}^N g_j(w)x^j\bigg|\sum_{i=1}^N g_i(w)(z_i^2-2\bar{z}z_i+\bar{z}^2)\\
		&\leq(\max_{i=1,\ldots,N}|x^i| + \sum_{j=1}^N g_j(w)|x^j| )\lf(\sum_{i=1}^N g_i(w)z_i^2-2\bar{z}\sum_{k=1}^N g_k(w)z_k+\bar{z}^2\rt)\\
		&=(\max_{i=1,\ldots,N}|x^i|+\max_{i=1,\ldots,N}|x^i|)\lf(\sum_{i=1}^N g_i(w)z_i^2-\bar{z}^2\rt)\\
		&\leq 2\max_{i=1,\ldots,N}|x^i| \lf(\sum_{i=1}^N z_i^2\rt)\\
		&\leq \mathcal{M}(x)|z|^2,
	\end{align*}
	where $\mathcal{M}(x):= 2\max_{i=1,\ldots,N}|x^i|$.
	Thus, we have, for all \(w,z\in \R^N\), that
	\[
	z^\top\nabla^2_w G(w,x)z + \mathcal{M}(x)|z|^2 \geq 0,
	\]
	which shows that, for each fixed $x \in \R^N$, $w \mapsto G(w,x)+\frac{\mathcal{M}(x)}{2}|w|^2$ is convex. By definition of semi-convexity, $w \mapsto G(w,x)$ is $\mathcal{M}(x)$-semi-convex. It follows that $\Th \mapsto G(w,x)-\eta$ is $\mathcal{M}(x)$-semi-convex, which completes the proof.
	\\
	
	\noindent\textit{Proof of Claim 2.} To prove $R(\Th,x)+\mathcal{M}(x)\Th \in \partial_\Th \Phi(\Th,x)$ for all $\Th \in \R^{N+1}$, we consider the following three cases:
	\begin{enumerate}
		\item When $F(\Th,x)>0$, i.e., $G(w,x)>\eta$, we have $I(\Th,x)=1$, $R(\Th,x)=\nabla_\Th F(\Th,x)$, and $\Phi(\Th,x)=F(\Th,x)+\frac{\mathcal{M}(x)}{2}|\Th|^2$. Since $\Phi(\Th,x)$ is differentiable in $\Th$ for $F(\Th,x)>0$, we get 
		\begin{equation}\label{1}
			\nabla_\Th \Phi(\Th,x) = \nabla_\Th F(\Th,x) +\mathcal{M}(x)\Th = R(\Th,x)+\mathcal{M}(x)\Th \in \partial_\Th \Phi(\Th,x).
		\end{equation}
		\item If $F(\Th,x)=0$, we have $I(\Th,x)=1$, $R(\Th,x)=\nabla_\Th F(\Th,x)$. Since $\Phi(\Th,x)$ may not be differentiable when $F(\Th,x)=0$, at the points where $F(\Th,x)=0$, we have
		\begin{equation*}
			\partial_\Th \Phi(\Th,x) = \text{conv}\{\nabla_\Th F(\Th,x)+\mathcal{M}(x)\Th,\;\mathcal{M}(x)\Th\}=\{\lbd R(\Th,x)+\mathcal{M}(x)\Th: \lbd\in[0,1]\}.
		\end{equation*}
		Letting $\lbd = 1$ yields
		\begin{equation}\label{2}
			R(\Th,x)+\mathcal{M}(x)\Th \in  \partial_\Th \Phi(\Th,x).
		\end{equation}
		\item For $F(\Th,x)<0$, we have $I(\Th,x)=0$, $R(\Th,x)=\mathbf{0} \in \R^{N+1}$, and $\Phi(\Th,x)=\frac{\mathcal{M}(x)}{2}|\Th|^2$. Since $\Phi(\Th,x)$ is differentiable in $\Th$ for $F(\Th,x)<0$, we obtain 
		\begin{equation}\label{3}
			\nabla_\Th \Phi(\Th,x) = \mathcal{M}(x)\Th = R(\Th,x)+\mathcal{M}(x)\Th \in \partial_\Th \Phi(\Th,x).
		\end{equation}
	\end{enumerate}
	Combining \eqref{1}--\eqref{3} yields $R(\Th,x)+\mathcal{M}(x)\Th \in \partial_\Th \Phi(\Th,x)$ for all $\Th \in \R^{N+1}$ and $x\in\R^N$.
\end{proof}

\begin{lemma}\label{semiconvex}
	Suppose Assumption~\ref{A1} holds. Then, the objective function $V$ defined in \eqref{CVaRproblem} is $k$-semi-convex, that is, $V+\frac{k}{2}|\cdot|^2$ is convex.
\end{lemma}
	\begin{proof}
	Define $\tl V(\Th):=	V(\Th)+\frac{k}{2}|\Th|^2$, then the gradient of $\tl{V}$ is given by $\nabla \tl V(\Th)=h(\Th)+k\Th$. By Lemma~\ref{A4}, we have, for any \(\Th,\Th'\in\R^{N+1}\), that
		\begin{align*}
		\langle
			\nabla \tl V(\Th)
			-\nabla \tl V(\Th'),
			\Th-\Th'
			\rangle & =
		\langle
			h(\Th)-h(\Th'),\Th-\Th'
		\rangle
			+k|\Th-\Th'|^2 \\
			&\quad \geq
			-k|\Th-\Th'|^2+k|\Th-\Th'|^2
			=0,
		\end{align*}
		Thus, \(\nabla\tl  V\) is monotone. Since \(\tl V\) is
		continuously differentiable on the convex set \(\R^{N+1}\), it follows that
		\(\tl V\) is convex. Consequently, \(V\) is \(k\)-semi-convex.
	\end{proof}

\section{Auxiliary Processes}
Consider the $\R^{N+1}$-valued overdamped Langevin SDE $(Z_t)_{t\geq0}$ given by
\begin{equation}\label{SDE}
	\rd Z_t = -h(Z_t)\rd t+\sqrt{2\beta^{-1}}\rd B_t ,\quad t\geq0, 
\end{equation}
with initial condition $Z_0:=\Th_0$, where $h = \nabla V$ and $(B_t)_{t\geq0}$ is a standard $(N+1)$-dimensional Brownian motion with its completed natural filtration $(\cF_t)_{t\geq0}$. Let $(P_t)_{t\geq0}$ denote the Markov semigroup associated with \eqref{SDE}.

We next introduce several auxiliary processes that will be used in the convergence analysis. For each $\lbd>0$, define the time-scaled process $(Z^\lbd_t)_{t\geq0}$ by $Z^{\lbd}_t := Z_{\lbd t}$, $t\geq0$. Then $(Z_t^\lbd)_{t \geq 0}$ satisfies
\begin{equation}\label{Zlbd}
	\rd Z^{\lbd}_t= -\lbd h (Z^{\lbd}_t)\rd t+\sqrt{2\lbd\beta^{-1}}\rd \tl{B}^{\lbd}_t ,\quad Z^{\lbd}_0=\Th_0, 
\end{equation}
where $\tl{B}^{\lbd}_t:=B_{\lbd t}/\sqrt{\lbd}$, $t\geq0$, is a standard Brownian motion with respect to the filtration $\cF_t^{\lbd}:=\cF_{\lbd t}$, $t\geq0$.

Next, we define the continuous-time interpolation of the SGLD algorithm \eqref{SGLD} by
\begin{equation}\label{barTh}
	\rd \bar{\Th}_t^\lbd =-\lbd H(\bar{\Th}_{\ftf}^\lbd,X_{\ctc})\rd t+\sqrt{2\lbd\beta^{-1}}\rd \tl{B}^{\lbd}_t ,\quad \bar{\Th}^\lbd_0=\Th_0.
\end{equation}
Then, at the grid points, the interpolated process has the same law as the SGLD iterates; that is, $\cL(\bar{\Th}_n^\lbd)=\cL(\Th_n^\lbd)$ for each $n\in\N_0$. 

\begin{definition}\label{def1}
Fix $n_0 \in\N_0$. For each $n \geq n_0$, let $\big(W^{\lbd,n}_t\big)_{t\in[0,n_0]}$ denote the Brownian increments $W^{\lbd,n}_t:=\tl{B}^{\lbd}_{t+(n-n_0)}-\tl{B}^{\lbd}_{(n-n_0)}$. Then, we define the restarted continuous-time process $(\zeta_t^{\lbd,n})_{t\in[0,n_0]}$ given by
\begin{equation}\label{zeta}
 \rd 	\bar{\zeta}_t^{\lbd,n} = -\lbd h(\bar{\zeta}_t^{\lbd,n})\rd t +\sqrt{2\lbd\beta^{-1}} \rd  W^{\lbd,n}_t,\quad 	\bar{\zeta}_0^{\lbd,n}=\Th^{\lbd}_{n-n_0},
\end{equation}
and the restarted interpolation process $(\vartheta^{\lbd,n}_t)_{t\in[0,n_0]}$ is defined by
\begin{equation}\label{vartheta}
 \rd \vartheta_t^{\lbd,n} = -\lbd H(\vartheta_{\ftf}^{\lbd,n},X_{\ctc})\rd t +\sqrt{2\lbd\beta^{-1}} \rd  W^{\lbd,n}_t,\quad 	\vartheta_0^{\lbd,n}=\Th^{\lbd}_{n-n_0}.
\end{equation}
\end{definition}

Note that, both processes share the same initial condition and driving noise, and evolve on a common time interval $[0,n_0]$.The restarted interpolation process $\vartheta^{\lbd,n}_t$ cast $n_0$ steps of SGLD \eqref{SGLD} which implies that $\cL(\vartheta^{\lbd,n}_{n_0})=\cL(\bar{\Th}^{\lbd}_{n})$. Moreover, the process $\bar{\zeta}^{\lbd,n}_{n_0}$ has the same law as the time-scaled Langevin SDE \eqref{Zlbd}, started at time $n-n_0$, with initial condition $\Th_{n - n_0}^\lbd$, and evolves over a fixed time interval of length $n_0$. This implies that $\cL(\bar{\zeta}_{n_0}^{\lbd,n}) = \cL(\bar{\Th}_{n - n_0}^\lbd)P_{n_0\lbd}$.

\section{Existence and uniqueness of solution to SDE and invariant measure}
Consider the infinitesimal generator $\mathscr G$ associated with the SDE \eqref{SDE}
defined for all $\phi \in C^2(\mathbb{R}^{N+1})$ and $\Th \in \R^{N+1}$ by
\begin{equation}\label{generator}
\mathscr G\phi(\Th)
=
-\langle h(\Th), \nabla \phi(\Th)\rangle
+
\beta^{-1}\Delta \phi(\Th).\tag{$\mathscr G$}
\end{equation}
Set the Lyapunov function $\mathcal{V}(\Th):=1+|\Th|^2$ for all $\Th\in\mathbb{R}^{N+1}$, then $\mathcal V\in C^2(\R^{N+1})$. By Lemma~\ref{A2} and Young's inequality, we obtain
\begin{align}
\mathscr{G} \mathcal{V}(\Th) =-\langle h(\Th), 2\Th\rangle
	+
	2(N+1)/\beta& \le  |h(\Th)| |2\Th|
	+
	2(N+1)/\beta \nonumber\\
	& \le  2\sqrt{m}|\Th| +2\sqrt{L}|\Th|^2
	+
	2(N+1)/\beta  \nonumber \\
	& \le \frac{m}{2\sqrt{L}} +4\sqrt{L}|\Th|^2
	+
    2(N+1) /\beta \nonumber \\
	& \le C_*\mathcal{V}(\Th), \label{17}
\end{align}
where $C_*=4\sqrt{L}+m/(2\sqrt{L})+2(N+1)/\beta$. Moreover, by Lemma~\ref{dissipativity},
\begin{equation}\label{GVx}
	\mathscr{G} \mathcal{V}(\Th)
	\le -\gm|\Th|^2+2b+2(N+1)/\beta =
	-\gm \mathcal{V}(\Th)+\gm+2b+2(N+1)/\beta.
\end{equation}
It follows that
 \begin{equation}\label{19}
 \lim_{|\Th|\to\infty} \mathcal{V}(\Th) = \infty,
\qquad
\lim_{|\Th|\to\infty} \mathscr{G} \mathcal{V}(\Th)  = -\infty. 
\end{equation}

\begin{proposition}
	Let Assumption \ref{A1} hold. Then the SDE \eqref{SDE} admits a unique strong solution and a unique invariant measure.
\end{proposition}

\begin{proof}
We first establish existence and uniqueness of the solution. To this end, we verify the
assumptions of \cite[Theorem 2.8]{GyongyKrylov}. By the linear growth condition in Lemma \ref{A2} and the fact that the diffusion coefficient is constant, together with the Lyapunov estimate~\eqref{17}, conditions (i),
(ii), and (iv) of \cite[Theorem 2.8]{GyongyKrylov} are satisfied. Moreover, condition
(iii) is immediate, since in the present setting the state space is
$D=\mathbb{R}^{N+1}$. Consequently, the SDE \eqref{SDE} admits a unique strong solution.

We next prove the existence of an invariant measure. For this purpose, we verify the assumptions of \cite[Theorem 2.2]{exists}. By Lemma~\ref{A2}, the drift coefficient is locally bounded and hence locally integrable. Thus, condition (A2) holds. Moreover, since the diffusion matrix is
constant and uniformly non-degenerate, condition (A1) is satisfied. Finally, the Lyapunov function
$\mathcal V$ satisfies \eqref{19}. Therefore, by \cite[Theorem 2.2]{exists}, there exists at least one invariant measure of the SDE \eqref{SDE}. Moreover, \eqref{GVx} implies that any invariant probability
measure has finite second moment, and hence belongs to $\mathcal{P}_2(\R^{N+1})$. Finally, the uniqueness of the invariant measure follows from Proposition~\ref{prop2} below, by the exponential Wasserstein contraction established there.
\end{proof}

\section{Contraction Estimates}\label{Contraction}
\begin{proposition}[\cite{lytras2026tamed}, Proposition 4]\label{propLSI}
Let Assumption \ref{A1} hold. Then, the invariant measure $\pi_\beta \propto e^{-\beta V}$ of SDE \eqref{SDE} satisfies a logarithmic Sobolev inequality with some constant $\rho(\beta)>0$. More precisely, for every $\phi \in C^1_{b}(\R^{N+1})$ with $\int_{\R^{N+1}} \phi^2 \rd \pi_\beta = 1$, we have
\begin{equation}\label{LSI}
\int_{\R^{N+1}} \phi^2 \log \phi^2 \rd \pi_\beta \leq \rho(\beta) \int_{\R^{N+1}} |\nabla \phi|^2 \rd \pi_\beta.
\end{equation}

\end{proposition}
\begin{proof} 
By Lemma~\ref{semiconvex}, the potential $V$ is $k$-semi-convex.  Since $\beta>0$, it follows that $\beta V$ is $\beta k$-semi-convex, or equivalently, $\nabla^2(\beta V)\succeq -\beta k I_{N+1}$ in the distributional sense. Moreover, Lemma~\ref{A3} establishes the required convexity-at-infinity condition. Hence, under Assumption~\ref{A1}, the assumptions of \cite[Proposition~4]{lytras2026tamed} are satisfied. Therefore, $\pi_\beta$ satisfies the logarithmic Sobolev inequality \eqref{LSI} for some constant $\rho(\beta)>0$. 
\end{proof}

\begin{proposition}[\cite{lytras2026tamed}, Proposition 5]\label{prop2}
Let Assumption \ref{A1} hold. Then, the logarithmic Sobolev inequality in \eqref{LSI} holds with \begin{equation}\label{LSIconstant}
\rho(\beta) \leq \frac{2}{\beta \gm} e^{\beta (4r(\sqrt{m}+\sqrt{L}r)+\gm r^2/2)}.\end{equation}
Moreover, $\pi_\beta$ is the unique invariant measure of \eqref{SDE} and, for all $t\geq0$ and $\nu \in \mathcal{P}_2(\R^{N+1})$,
	\begin{equation}\label{contractionr}
	W_2(\nu P_t, \pi_\beta) \leq C_{w}e^{-C_{r}t} W_2(\nu,\pi_\beta),
	\end{equation}
	where, with $t_* = 1/(2k)\log\big(1+\beta k \rho(\beta)/2\big)$, $C_{r} = 2/(\beta \rho(\beta))$ and
\[
	C_{w} =\max\lf\{e^{(k+2/(\beta\rho(\beta)))t_*},e^{2t_*/(\beta \rho(\beta))}\sqrt{\frac{\beta k\rho(\beta)}{2(1-e^{-2kt_*})}} \rt\} \leq \sqrt{e\bigg(1+\frac{\beta k \rho(\beta)}{2}\bigg)}.
\]
The constants $m,L,r,k$ are given in Table \ref{table1}.
\end{proposition}
\begin{proof}
Under Assumption \ref{A1}, Lemmas \ref{A2}, \ref{A3}, \ref{A4}, and \ref{semiconvex} ensure that all the assumptions of \cite[Remark 10 and Proposition 5]{lytras2026tamed} are satisfied. Therefore, by \cite[Remark 10]{lytras2026tamed}, the logarithmic Sobolev inequality \eqref{LSI} holds with a constant $\rho(\beta)$ satisfying \eqref{LSIconstant}, and
the exponential contraction estimate \eqref{contractionr} follows
directly from \cite[Proposition 5]{lytras2026tamed}. Finally, the
contraction estimate implies the uniqueness of the invariant measure.
\end{proof}

\begin{remark}\label{remarkn0}
    For the remainder of the analysis, we fix the restart horizon in Definition \ref{def1} as
    \begin{equation}\label{n0}
        n_0:= \lf\lceil \frac{\ln C_w +1}{C_r \lbd} \rt\rceil, \quad \text{so that} \quad \lbd n_0 \leq  \frac{\ln C_w +1}{C_r}+\lbd_0 =:T_0,
    \end{equation}
    where $\lbd \in (0,\lbd_0)$ with $\lbd_0=\min\{1,1/\gm,\gm/L\}$, $C_w$ and $C_r$ are the contraction constants of Proposition \ref{prop2}.
\end{remark}

\section{Preliminary Estimates}
\begin{lemma}\label{C1}
Let Assumption \ref{A1} hold. Then
\[
\sup_{t \geq 0 }\E|Z_t|^2 \leq \E|\Th_0|^2+C_1,
\]
where $C_1 = (4/\gm)(b+(N+1)/\beta)$.
\end{lemma}
\begin{proof}
For $\mathfrak{r}>0$, define the stopping time $\tau_{\mathfrak{r}} = \inf\{t\geq0:|Z_t|\geq \mathfrak{r}\}$. By It\^{o}'s formula, we have, for all $t\geq0$, that
	\begin{align*}
	e^{\gm(t\wedge \tau_\mathfrak{r})/2}|Z_{t\wedge\tau_\mathfrak{r}}|^2 &= |\Th_0|^2  + \int_0^{t\wedge \tau_\mathfrak{r}}
		e^{\gm s/2}
		\left[
		\frac{\gm}{2}|Z_s|^2
		-2\langle Z_s,h(Z_s)\rangle
		+\frac{2(N+1)}{\beta}
		\right]\rd s \\
		&\quad + \sqrt{8\beta^{-1}}
		\int_0^{t\wedge \tau_\mathfrak{r}}
		e^{\gm s/2}
		\left\langle Z_s,\rd B_s\right\rangle .
	\end{align*}
	Since the stochastic integral is a true martingale, taking expectations on both sides and applying Lemma \ref{dissipativity} yield
	\begin{align*}
		\E\lf[e^{\gm(t\wedge \tau_\mathfrak{r})/2}|Z_{t\wedge\tau_\mathfrak{r}}|^2 \rt]&= \mathbb{E}|\Th_0|^2 + \mathbb{E}\int_0^{t\wedge \tau_\mathfrak{r}}
		e^{\gm s/2}
		\left[
		\frac{\gm}{2}|Z_s|^2
		-2\langle Z_s,h(Z_s)\rangle
		+\frac{2(N+1)}{\beta}
		\right]\rd s\\
		&\leq \mathbb{E}|\Th_0|^2 + \mathbb{E}\int_0^{t\wedge \tau_\mathfrak{r}}
		e^{\gm s/2}
		\left[
		-\frac{\gm}{2}|Z_s|^2
		+2\lf(b+\frac{N+1}{\beta}\rt)
		\right]\rd s\\
		&= \mathbb{E}|\Th_0|^2 
		+\frac{4}{\gm}\lf(b+\frac{N+1}{\beta}\rt)\E[e^{\gm(t\wedge \tau_\mathfrak{r}) /2}]
		- 	\frac{\gm}{2} \E \int_0^{t\wedge \tau_\mathfrak{r}}
		e^{\gm s/2}
	|Z_s|^2
		\rd s\\
		&\leq  \E |\Th_0|^2 
		+\frac{4}{\gm}\lf(b+\frac{N+1}{\beta}\rt)e^{\gm t /2}.
	\end{align*}
Hence, by Fatou's lemma, we obtain
\begin{align*}
		e^{\gm t/2}\E|Z_{t}|^2 &= \E\lf[\liminf_{\mathfrak{r}\rightarrow \infty} e^{\gm(t\wedge \tau_\mathfrak{r})/2}|Z_{t\wedge\tau_\mathfrak{r}}|^2\rt]\leq \liminf_{\mathfrak{r}\rightarrow \infty}\E\lf[ e^{\gm(t\wedge \tau_\mathfrak{r})/2}|Z_{t\wedge\tau_\mathfrak{r}}|^2\rt]\\
		&\leq \E |\Th_0|^2 
		+\frac{4}{\gm}\lf(b+\frac{N+1}{\beta}\rt)e^{\gm t /2}.
\end{align*}
Therefore,
\[
\E|Z_{t}|^2 \leq e^{-\gm t/2}\E|\Th_0|^2
+\frac{4}{\gm}\lf(b+\frac{N+1}{\beta}\rt).
\]
Taking the supremum over $t \geq 0$ completes the proof.
\end{proof}

\begin{lemma}\label{C2}
Suppose Assumption \ref{A1} holds and let $\lbd_0=\min\{1,1/\gm,\gm/L\}$. For every $\lbd \in (0,\lbd_0)$,
\[
\sup_{n\in \N_0} \E|\Th_n^\lbd|^2 \leq \E|\Th_0|^2+C_2,
\]
where $C_2= (2b+\gm m/L + 2(N+1)/\beta)/(\gm - \lbd L)$.
\end{lemma}
\begin{proof}
	By the definition of SGLD algorithm in \eqref{SGLD}, we have, for all $n \in \N_0$, that
	\[
	|\Th_{n+1}^\lbd|^2=|\Th^\lbd_{n}-\lbd H(\Th^\lbd_{n},X_{n+1})|^2+\frac{2\lbd}{\beta}|\xi_{n+1}|^2+2\sqrt{2\lbd \beta^{-1}}\langle \Th^{\lbd}_n - \lbd H(\Th^{\lbd}_n ,X_{n+1}),\xi_{n+1} \rangle.
	\]
Since $\xi_{n+1}$ is independent of $\Th_n^{\lbd}$ and $X_{n+1}$, taking conditional expectation on both sides yields
\[
\E[|\Th_{n+1}^{\lbd}|^2 \vert \Th_{n}^{\lbd}] = \E[|\Th_{n}^{\lbd}|^2 \vert \Th_{n}^{\lbd}] -2\lbd \E[\langle\Th^\lbd_{n}, H(\Th^\lbd_{n},X_{n+1}) \rangle \vert  \Th_{n}^{\lbd}]+\lbd^2\E[ |H(\Th^\lbd_{n},X_{n+1})|^2 \vert  \Th_{n}^{\lbd}] +\frac{2\lbd(N+1)}{\beta}.
\]
Using Lemmas \ref{A2} and \ref{dissipativity}, we obtain
\begin{align*}
\E[|\Th_{n+1}^{\lbd}|^2 \vert \Th_{n}^{\lbd}] &\leq |\Th_{n}^{\lbd}|^2 - 2\lbd \lf(\frac{\gm}{2}|\Th_{n}^{\lbd}|^2-b\rt)+\lbd^2(m+L|\Th_{n}^{\lbd}|^2)+\frac{2\lbd(N+1)}{\beta}\\
&\leq (1-\lbd \gm+\lbd^2 L)|\Th_{n}^{\lbd}|^2+\lbd \lf(2b+m\lbd+\frac{2(N+1)}{\beta}\rt).
\end{align*}
Taking expectation gives
\[
\E|\Th_{n+1}^{\lbd}|^2 \leq (1-\lbd \gm+\lbd^2 L)\E|\Th_{n}^{\lbd}|^2+\lbd \lf(2b+m\lbd+\frac{2(N+1)}{\beta}\rt).
\]
Since $\lbd\in(0,\min\{1/\gm,\gm/L\})$, we have $1-\lbd \gm+\lbd^2 L \in(0,1)$. Iterating the above inequality yields
\begin{align}\label{D21}
	\E|\Th_{n}^{\lbd}|^2 
	&\leq (1-\lbd \gm+\lbd^2 L)^n  \E|\Th_{0}|^2+
	\frac{1-(1-\lbd \gm+\lbd^2 L)^n}{ \gm-\lbd L}
	 \lf(2b+\frac{\gm m}{L}+\frac{2(N+1)}{\beta}\rt)\nonumber\\
	&\leq \E|\Th_{0}|^2 +C_2,
\end{align}
where $C_2:= (2b+\gm m/L + 2(N+1)/\beta)/(\gm - \lbd L)>0$.
\end{proof}

\begin{lemma}\label{C3}
Let Assumption \ref{A1} hold and set $\lbd_0=\min\{1,1/\gm,\gm/L\}$. Then, for every $\lbd \in (0,\lbd_0)$,
\[
\sup_{t\geq0}\E|\bar{\Th}_{\ftf}^\lbd-\bar{\Th}_t^\lbd|^2 \leq \lbd\lf(2\gm\E|\Th_0|^2+C_3\rt),	
\]
where $C_3 =  2\gm C_2  + 2\gm m/L +4(N+1)/\beta$ with $C_2$ given in Lemma \ref{C2}.
\end{lemma}
\begin{proof}
By definition of $\bar{\Th}_t^\lbd$ in \eqref{barTh}, we obtain
\begin{align*}
\E |\bar{\Th}_{t}^\lbd-\bar{\Th}_{\lfloor t \rfloor}^\lbd|^2 &\leq 2 \lbd^2\E \lf| \int_{\ftf}^t H(\bar{\Th}_{\lfloor s \rfloor}^{\lbd},X_{\lceil s \rceil}) \rd s\rt|^2+\frac{4\lbd}{\beta} \E \lf| \tl{B}^\lbd_t - \tl{B}^\lbd_{\ftf} \rt|^2\\
&\leq 2 \lbd^2(t-\ftf)^2 \E\lf|H(\bar{\Th}_{\lfloor t \rfloor}^{\lbd},X_{\lceil t \rceil})\rt|^2+\frac{4\lbd(N+1)}{\beta}\\
&\leq  2 \lbd^2 \E\lf|H(\bar{\Th}_{\lfloor t \rfloor}^{\lbd},X_{\ftf +1})\rt|^2+\frac{4\lbd(N+1)}{\beta}.
\end{align*}
Applying Lemmas \ref{A2} and \ref{C2} yields
\begin{align*}
	\E |\bar{\Th}_{t}^\lbd-\bar{\Th}_{\lfloor t \rfloor}^\lbd|^2 &\leq  2 \lbd^2 \lf(m+L\E|\bar{\Th}_{\lfloor t \rfloor}^{\lbd}|^2\rt)+\frac{4\lbd(N+1)}{\beta}\\
	&\leq 2 \lbd^2L\lf(  \E|\Th_0|^2+C_2   \rt) + 2 \lbd^2 m +\frac{4\lbd(N+1)}{\beta}\\
	&\leq \lbd \lf(   2\lbd L \E|\Th_0|^2+2\lbd L C_2 + 2 \lbd m +\frac{4(N+1)}{\beta} \rt).
\end{align*}
Since $\lbd<\gm/L$, it follows that
\[
\E |\bar{\Th}_{t}^\lbd-\bar{\Th}_{\lfloor t \rfloor}^\lbd|^2\leq \lbd \lf( 2\gm \E|\Th_0|^2+ 2\gm C_2  + \frac{2\gm m}{L} +\frac{4(N+1)}{\beta} \rt)\leq  \lbd\lf( 2\gm \E|\Th_0|^2+ C_3 \rt),
\]
where $C_3 =  2\gm C_2  + 2\gm m/L +4(N+1)/\beta$. This completes the proof.
\end{proof}

\begin{lemma}\label{C4}
Let Assumption \ref{A1} hold and set $\lbd_0=\min\{1,1/\gm,\gm/L\}$. Then, for every $\lbd \in (0,\lbd_0)$, and $n\geq n_0$ with $n_0$ defined in \eqref{n0},
\[
\sup_{t\in[0,n_0]} \E|\bar{\zeta}^{\lbd,n}_t|^2 \leq \E|\Th_0|^2+C_4,	
\]
where $C_4 =C_2+2T_0 \lf((N+1)/\beta +b\rt)$ with $C_2$ given in Lemma \ref{C2} and $T_0$ given in \eqref{n0}.
\end{lemma}
\begin{proof}
	Applying Itô's formula to $ |\bar{\zeta}_t^{\lbd,\mathfrak{n}}|^2$ yields
	\begin{align*}
		|\bar{\zeta}_t^{\lbd,n}|^2 
		&=
		|\bar{\Th}_{n-n_0}^{\lbd}|^2
		+
		\int_{0}^{t}
		\left[
		-2\lbd
		\left\langle
		\bar{\zeta}_s^{\lbd,n},
		h(\bar{\zeta}_s^{\lbd,n})
		\right\rangle
		+
		\frac{2\lbd (N+1)}{\beta}
		\right]\rd s
		+
		2\sqrt{2\lbd\beta^{-1}}
		\int_{0}^{t}
		\left\langle
		\bar{\zeta}_s^{\lbd,n},
		\rd W_s^{\lbd,n}
		\right\rangle.
	\end{align*}
By a standard stopping-time argument, the stochastic integral is a martingale with zero expectation. Hence, taking expectations and using Lemma~\ref{dissipativity}, we obtain
		\begin{align*}
		\E|\bar{\zeta}_t^{\lbd,\mathfrak{n}}|^2
		&=
		\E|\bar{\Th}_{n-n_0}^{\lbd}|^2
		+
		\int_{0}^{t}
		\left[
		-2\lbd
		\E\left\langle
		\bar{\zeta}_s^{\lbd,n},
		h(\bar{\zeta}_s^{\lbd,n})
		\right\rangle
		+
		\frac{2\lbd (N+1)}{\beta}
		\right]\rd s\\
		&\leq \E|\bar{\Th}_{n-n_0}^{\lbd}|^2
		+
		\int_{0}^{t}
		\left[
		-2\lbd\lf(\frac{\gm}{2}\E|\bar{\zeta}_s^{\lbd,n}|^2 -b \rt)+
		\frac{2\lbd (N+1)}{\beta}
		\right]\rd s.
	\end{align*}
Differentiating both sides and multiplying by $e^{\lbd\gm t}$ yield
\[
\frac{\rd}{\rd t} e^{\lbd \gm t}\E|\bar{\zeta}_t^{\lbd,n}|^2 \leq 2\lbd \lf(\frac{N+1}{\beta} +b\rt)e^{\lbd \gm t},
\]
which implies, by integrating over
$[0,t]$, that
\begin{align*}
	\E|\bar{\zeta}_t^{\lbd,n}|^2 &\leq e^{-\lbd \gm t}\E|
	\bar{\Th}_{n-n_0}^{\lbd}|^2+2\lbd \lf(\frac{N+1}{\beta} +b\rt)\frac{1-e^{-\lbd \gm t}}{\lbd \gm}\\
	&\leq e^{-\lbd \gm t}\E|
	\bar{\Th}_{n-n_0}^{\lbd}|^2+2\lbd \lf(\frac{N+1}{\beta} +b\rt)t.
\end{align*}
Therefore, by Lemma~\ref{C2}, we have
\[
\sup_{t\in[0,n_0]}\E|\bar{\zeta}_t^{\lbd,n}|^2 \leq \E|\Th_0|^2+C_4,
\]
where $C_4:=C_2+2 T_0\lf((N+1)/\beta +b\rt)$.
\end{proof}

\begin{lemma}\label{C5}
Let Assumption \ref{A1} hold and set $\lbd_0=\min\{1,1/\gm,\gm/L\}$. Then, for every $\lbd \in (0,\lbd_0)$, $t\in[0,n_0]$, and $n \geq n_0$ with $n_0$ defined in \eqref{n0}
\[
W_2\lf( \cL(\vartheta_t^{\lbd,n}),\cL(\bar{\zeta}^{\lbd,n}_t) \rt)\leq C_5\lbd^{1/4},	
\]
where 
\begin{align*}
		C_5&=\sqrt{2T_0}e^{2kT_0} \lf(  2\gm \E|\Th_0|^2+C_3\rt)^{1/4} \\
        &\quad \times\lf(2k\lf( 2\gm \E|\Th_0|^2+C_3\rt)^{1/2} +	\lf(4m+4L\E|\Th_0|^2+2L(C_2
		+C_4)\rt)^{1/2}\rt)^{1/2},
\end{align*}
with $C_2$ given in Lemma \ref{C2}, $C_3$ given in Lemma \ref{C3}, $C_4$ given in Lemma \ref{C4}, and $T_0$ given in \eqref{n0}.
\end{lemma}
\begin{proof}
Recall $(\vartheta_t^{\lbd,n})_{t\in[0,n_0]}$ and $(\bar{\zeta}^{\lbd,n}_t )_{t\in[0,n_0]}$ defined in \eqref{vartheta} and \eqref{zeta}, respectively. Applying It\^o's formula to $|\vartheta_t^{\lbd,n}-\bar{\zeta}^{\lbd,n}_t |^2$ and taking expectation yield
\[
\E| \vartheta_t^{\lbd,n}-\bar{\zeta}^{\lbd,n}_t |^2  =-2\lbd
\E\int_{0}^{t}
\left\langle
 \vartheta_s^{\lbd,n}-\bar{\zeta}^{\lbd,n}_s,
H\left(\vartheta_{\lfloor s\rfloor}^{\lbd,n},X_{\lceil s\rceil}\right)
-
h\left(\bar{\zeta}_s^{\lbd,n}\right)\right\rangle
\rd s = A_1(t)+A_2(t),
\]
where
\begin{align*}
&A_1(t):=-2\lbd
\E\int_{0}^{t}
\left\langle
\vartheta_s^{\lbd,n}-\vartheta_{\lfloor s\rfloor}^{\lbd,n},
H\left(\vartheta_{\lfloor s\rfloor}^{\lbd,n},X_{\lceil s\rceil}\right)
-
h\left(\bar{\zeta}_s^{\lbd,n}\right)\right\rangle
\rd s,\\
&A_2(t):= -2\lbd
\E\int_{0}^{t}
\left\langle
\vartheta_{\lfloor s\rfloor}^{\lbd,n}-\bar{\zeta}_s^{\lbd,n},
H\left(\vartheta_{\lfloor s\rfloor}^{\lbd,n},X_{\lceil s\rceil}\right)
-
h\left(\bar{\zeta}_s^{\lbd,n}\right)\right\rangle
\rd s.
\end{align*}
We first estimate $A_1(t)$. By Cauchy-Schwarz inequality and Lemma \ref{A2}, we have
\begin{align*}
	A_1(t)&\leq 2\lbd
	\int_{0}^{t} \lf(\E\lf|
	\vartheta_s^{\lbd,n}-\vartheta_{\lfloor s\rfloor}^{\lbd,n}\rt|^2\rt)^{1/2}
	\lf(\E\lf|H\left(\vartheta_{\lfloor s\rfloor}^{\lbd,n},X_{\lceil s\rceil}\right)
-
h\left(\bar{\zeta}_s^{\lbd,n}\right)\rt|^2\rt)^{1/2}
	\rd s \nonumber \\
	&\leq 2\lbd
	\int_{0}^{t} \lf(\E\lf|
	\vartheta_s^{\lbd,n}-\vartheta_{\lfloor s\rfloor}^{\lbd,n}\rt|^2\rt)^{1/2}
	\lf(2\E\lf|H\left(\vartheta_{\lfloor s\rfloor}^{\lbd,n},X_{\lceil s\rceil}\right)\rt|^2
	+2\E\lf|
	h\left(\bar{\zeta}_s^{\lbd,n}\right)\rt|^2\rt)^{1/2}
	\rd s \nonumber\\
		&\leq 2\lbd
	\int_{0}^{t} \lf(\E\lf|
	\vartheta_s^{\lbd,n}-\vartheta_{\lfloor s\rfloor}^{\lbd,n}\rt|^2\rt)^{1/2}
	\lf(4m+2L\lf(\E\lf|\vartheta_{\lfloor s\rfloor}^{\lbd,n}\rt|^2
	+\E\lf|
	\bar{\zeta}_s^{\lbd,n}\rt|^2\rt)\rt)^{1/2}
	\rd s.
\end{align*}
We then use the estimates in Lemmas \ref{C2}--\ref{C4} to obtain
\begin{align}\label{A_1}
	A_1(t)
    &\leq 2 \lbd^{3/2} \lf(  2\gm \E|\Th_0|^2+C_3\rt)^{1/2} 	\int_{0}^{n_0} \lf(4m+2L\lf(2\E|\Th_0|^2+C_2
	+C_4\rt)\rt)^{1/2}\nonumber\\
		&\leq 2 \lbd^{3/2} \lf(  2\gm\E|\Th_0|^2+C_3\rt)^{1/2} 	\lf(4m+4L\E|\Th_0|^2+2L(C_2
	+C_4)\rt)^{1/2}n_0.
\end{align}
Next, we estimate $A_2(t)$. We rewrite $A_2(t)$ as follows:
\begin{align*}
	A_2(t)=& -2\lbd
\E\int_{0}^{t}
\left\langle
\vartheta_{\lfloor s\rfloor}^{\lbd,n}-\bar{\zeta}_s^{\lbd,n},
h\left(\vartheta_{\lfloor s\rfloor}^{\lbd,n}\right)
-
h\left(\bar{\zeta}_s^{\lbd,n}\right)\right\rangle
\rd s,\\
&-2\lbd
\E\int_{0}^{t}
\left\langle
\vartheta_{\lfloor s\rfloor}^{\lbd,n}-\bar{\zeta}_s^{\lbd,n},
H\left(\vartheta_{\lfloor s\rfloor}^{\lbd,n},X_{\lceil s\rceil}\right)
-
h\left(\vartheta_{\lfloor s\rfloor}^{\lbd,n}\right)\right\rangle
\rd s.
\end{align*}
For $s\geq0$, define
$\mathcal H_s:=\cF_\infty^\lbd\vee\sg(\Th_0)\vee\mathcal G_{\lfloor s\rfloor}.$
Then $\vartheta_{\lfloor s\rfloor}^{\lbd,n}$ and
$\bar{\zeta}_s^{\lbd,n}$ are $\mathcal H_s$-measurable, while
$X_{\lceil s\rceil}$ is independent of $\mathcal H_s$. Hence, it holds almost surely that,
\[
\E\left[
H\left(\vartheta_{\lfloor s\rfloor}^{\lbd,n},X_{\lceil s\rceil}\right)
-
h\left(\vartheta_{\lfloor s\rfloor}^{\lbd,n}\right)
\bigg|
\mathcal H_s
\right]=0.
\]
Using this identity, Lemmas~\ref{A4} and \ref{C3}, we obtain
\begin{align}\label{A_2}
	A_2(t)&\leq -2\lbd
	\int_{0}^{t}
	\E\lf[\E\lf[\left\langle
	\vartheta_{\lfloor s\rfloor}^{\lbd,n}-\bar{\zeta}_s^{\lbd,n},
	H\left(\vartheta_{\lfloor s\rfloor}^{\lbd,n},X_{\lceil s\rceil}\right)
	-
	h\left(\vartheta_{\lfloor s\rfloor}^{\lbd,n}\right)\right\rangle \bigg\vert \mathcal{H}_s\rt]\rt]
	\rd s\nonumber\\
	&\quad +2k\lbd
	\int_{0}^{t} \E
	\lf| \vartheta_{\lfloor s\rfloor}^{\lbd,n}-\bar{\zeta}_s^{\lbd,n}\rt|^2
	\rd s\nonumber\\
	&= -2\lbd
	\int_{0}^{t}
	\E\lf[\left\langle
	\vartheta_{\lfloor s\rfloor}^{\lbd,n}-\bar{\zeta}_s^{\lbd,n},\E\lf[
	H\left(\vartheta_{\lfloor s\rfloor}^{\lbd,n},X_{\lceil s\rceil}\right)
	-
	h\left(\vartheta_{\lfloor s\rfloor}^{\lbd,n}\right)\bigg\vert \mathcal{H}_s\rt]\right\rangle \rt]
	\rd s\nonumber\\
	&\quad +4k\lbd
	\int_{0}^{t} \E
	\lf|\vartheta_{\lfloor s\rfloor}^{\lbd,n}-\vartheta_{s}^{\lbd,n}\rt|^2
	\rd s+4k\lbd
	\int_{0}^{t} \E
	\lf| \vartheta_s^{\lbd,n}-\bar{\zeta}_s^{\lbd,n}\rt|^2
	\rd s\nonumber\\
	&\leq 4k\lbd
	\int_{0}^{t} \E
	\lf|  \vartheta_s^{\lbd,n}-\bar{\zeta}_s^{\lbd,n}\rt|^2
	\rd s+4k \lbd^2 n_0 \lf(  2\gm\E|\Th_0|^2+C_3\rt).
\end{align}
Combining \eqref{A_1} and \eqref{A_2} yields
\begin{align*}
\E|  \vartheta_t^{\lbd,n}-\bar{\zeta}_t^{\lbd,n} |^2 &\leq 4k\lbd
	\int_{0}^{t} \E
	\lf|  \vartheta_s^{\lbd,n}-\bar{\zeta}_s^{\lbd,n}\rt|^2
\rd s+4k \lbd^2 n_0 \lf(  2\gm\E|\Th_0|^2+C_3\rt) \\
& \quad+ 2 \lbd^{3/2} \lf(  2\gm\E|\Th_0|^2+C_3\rt)^{1/2} 	\lf(4m+4L\E|\Th_0|^2+2L(C_2
	+C_4)\rt)^{1/2}n_0\\
&\leq 4k\lbd
	\int_{0}^{t} \E
	\lf|  \vartheta_s^{\lbd,n}-\bar{\zeta}_s^{\lbd,n}\rt|^2
\rd s+2 \lbd^{3/2} n_0 \bigg\{2k\lf(  2\gm\E|\Th_0|^2+C_3\rt) \\
& \quad+ \lf(  2\gm\E|\Th_0|^2+C_3\rt)^{1/2} 	\lf(4m+4L\E|\Th_0|^2+2L(C_2
	+C_4)\rt)^{1/2}\bigg\}<\infty
\end{align*}
Thus, applying Gr\"onwall's lemma on $[0,n_0]$ gives
\[
\E|  \vartheta_t^{\lbd,n}-\bar{\zeta}_t^{\lbd,n} |^2 \leq \lbd^{1/2} C e^{4k\lbd n_0} \leq \lbd^{1/2} Ce^{4kT_0}, 
\]
where
\begin{align*}
C = 2T_0\lf(  2\gm\E|\Th_0|^2+C_3\rt)^{1/2} \lf(2k\lf(  2\gm\E|\Th_0|^2+C_3\rt)^{1/2} +	\lf(4m+4L\E|\Th_0|^2+2L(C_2
+C_4)\rt)^{1/2}\rt).
\end{align*}
Finally, noticing $W_2\lf( \cL(\vartheta_t^{\lbd,n}),\cL(\bar{\zeta}^{\lbd,n}_t) \rt) \leq \sqrt{\E| \vartheta_t^{\lbd,n}-\bar{\zeta}^{\lbd,n}_t |^2 }$ completes the proof.
\end{proof}

\begin{lemma}\label{C6}
Let Assumption \ref{A1} hold and set $\lbd_0=\min\{1,1/\gm,\gm/L\}$. Then, for every $\lbd \in (0,\lbd_0)$ and $n \in N_0$,
\[
W_2\lf( \cL(\Th_n^{\lbd}),\pi_\beta \rt)\leq C_6,	
\]
where $C_6 :=  \lf(2(\E|\Th_0|^2+C_2) +2(\gm+2b+2(N+1)/\beta)/\gm\rt)^{1/2}$ with $C_2$ given in Lemma \ref{C2}.
\end{lemma}
\begin{proof}
Let $\Th_\infty \sim \pi_\beta$ be independent of $\Th_n^{\lbd}$. The independent coupling and Lemma \ref{C2} give that
\begin{align*}
W_2^2(\cL(\Th_n^{\lbd}),\pi_\beta) &\leq \E[|\Th^{\lbd}_n - \Th_{\infty}|^2]\\
&\leq 2\E[|\Th_n^{\lbd}|^2]+2\int_{\R^{N+1}}|x|^2\pi_\beta (\rd x) \\
&\leq 2(\E|\Th_0|^2+C_2)+2\int_{\R^{N+1}}\mathcal{V}(x)\pi_\beta (\rd x).
\end{align*}
Since $\pi_\beta$ is the invariant measure of SDE \eqref{SDE}, we have $\int_{\R^{N+1}} \mathscr G \mathcal{V}(x) \pi_\beta(\rd x) = 0$. Moreover, by the Lyapunov estimate~\eqref{GVx}, we have
\begin{align}\label{Vx1}
    \int_{\R^{N+1}} \mathcal{V}(x) \pi_\beta(\rd x) &\leq -\frac{1}{\gm}\int_{\R^{N+1}}\mathscr G \mathcal{V}(x) \pi_\beta(\rd x)
     +\frac{\gm+2b+2(N+1)/\beta}{\gm}\nonumber \\
     &\leq \frac{\gm+2b+2(N+1)/\beta}{\gm}.
\end{align}
Consequently,
\[
W_2(\cL(\Th_n^{\lbd}),\pi_\beta) \leq \lf(2(\E|\Th_0|^2+C_2) +2\frac{\gm+2b+2(N+1)/\beta}{\gm}\rt)^{1/2}.
\]
\end{proof}

\section{Estimates for the excess risk}
By using previous results, we can now provide an upper estimate for the expected excess risk associated with $V$ defined in \eqref{CVaRproblem}. To this end, we consider the following decomposition
\begin{equation}\label{EER}
\E[V(\Th^\lbd_n)]-\inf_{\Th \in \R^{N+1}}V(\Th)= \big(\E[V(\Th^\lbd_n)]-\E[V(\Th_\infty)]\big)+\big(\E[V(\Th_\infty)]-\inf_{\Th \in \R^{N+1}}V(\Th)\big).\tag{EER}
\end{equation}
In the following lemmas, we establish an error bound for each of the terms on the RHS of \eqref{EER}.
\begin{lemma}\label{CV1}
	Let Assumption \ref{A1} hold and set $\lbd_0=\min\{1,1/\gm,\gm/L\}$. Then, for every $\lbd \in (0,\lbd_0)$ and $n\in \N_0$, the following holds:
	\[
   \E[V(\Th^{\lbd}_n)]-\E[V(\Th_\infty)]  \leq C_{V}W_2(\cL(\Th^{\lbd}_n), \pi_\beta),	
	\]
	where $\Th_\infty \sim \pi_\beta$ with $\pi_\beta$ defined in \eqref{gibbs} and 
	\[
	C_{V}=\sqrt{m}+\frac{\sqrt{L}}{2}\sqrt{\E|\Th_0|^2+C_2}+\frac{\sqrt{L}}{2}\sqrt{\frac{\gm+2b+2(N+1)/\beta}{\gm}},
	\]
    with $C_2$ given in Lemma \ref{C2}.
\end{lemma}
\begin{proof}
	By Lemma~\ref{semiconvex}, $V$ is semi-convex and hence locally Lipschitz
	continuous \cite[Proposition 2.1]{alberti1992singularities}. Therefore, for any \(x,y\in\R^{N+1}\), the function $ g(t):=V(tx+(1-t)y)$, $\forall t\in[0,1]$,
	is Lipschitz continuous and hence absolutely continuous on \([0,1]\). Moreover, $g'(t) = \langle h(tx+(1-t)y),x-y\rangle$ for $t\in[0,1]$. Hence, by Lemma~\ref{A2}, we obtain
	\begin{equation}\label{L131}
	\begin{aligned}
	V(x)-V(y)=g(1)-g(0)&=\int_0^1 \langle h(tx+(1-t)y),x-y\rangle \rd t\\
	&\leq \int_0^1 |h(tx+(1-t)y)||x-y| \rd t\\
	&\leq \int_0^1 \lf(\sqrt{m}+\sqrt{L}|tx+(1-t)y|\rt)|x-y| \rd t\\
	&\leq  \lf(\sqrt{m}+\frac{\sqrt{L}}{2}|x|+\frac{\sqrt{L}}{2}|y|\rt)|x-y|.
	\end{aligned}
	\end{equation}
    Let $\cL(\Th_n^{\lbd})$ and $\cL(\Th_{\infty})=\pi_\beta$ follow an optimal coupling in $W_2$, i.e., $W_2^2(\cL(\Th_n^{\lbd}),\pi_\beta)=\E|\Th_n^{\lbd}-\Th_{\infty}|^2$. Taking expectations in \eqref{L131} and applying Cauchy-Schwarz and Minkowski's inequalities yield that
    \begin{align*}
      \E[V(\Th_n^{\lbd})-V(\Th_{\infty})] &\leq \lf(\E\lf[\lf(\sqrt{m}+\frac{\sqrt{L}}{2}|\Th_n^{\lbd}|+\frac{\sqrt{L}}{2}|\Th_{\infty}|\rt)^2\rt]\rt)^{1/2} \lf(\E|\Th_n^{\lbd}-\Th_{\infty}|^2\rt)^{1/2}\\
     &\leq \lf(\sqrt{m}+\frac{\sqrt{L}}{2}\sqrt{\E|\Th_n^{\lbd}|^2}+\frac{\sqrt{L}}{2}\sqrt{\E|\Th_{\infty}|^2}\rt)W_2(\cL(\Th_n^{\lbd}),\pi_\beta).
    \end{align*}
  We provide an upperbound for the second moment of $\pi_\beta$. Recall the generator $\mathscr{G}$ defined in \eqref{generator}. Since $\pi_\beta$ is the invariant measure of SDE \eqref{SDE}, we have $\int_{\R^{N+1}} \mathscr G \mathcal{V}(x) \pi_\beta(\rd x) = 0$. Moreover, by the Lyapunov estimate~\eqref{GVx}, we obtain
    \begin{align*}
    \E|\Th_\infty|^2 \leq \int_{\R^{N+1}} \mathcal{V}(x) \pi_\beta(\rd x) &\leq -\frac{1}{\gm}\int_{\R^{N+1}}\mathscr G \mathcal{V}(x) \pi_\beta(\rd x)
     +\frac{\gm+2b+2(N+1)/\beta}{\gm}\\
     &\leq \frac{\gm+2b+2(N+1)/\beta}{\gm}.
    \end{align*}
   Consequently,
      \[
    \E[V(\Th^\lbd_n)-V(\Th_\infty)]  \leq \lf(\sqrt{m}+\frac{\sqrt{L}}{2}\sqrt{\E|\Th_0|^2+C_2}+\frac{\sqrt{L}}{2}\sqrt{\frac{\gm+2b+2(N+1)/\beta}{\gm}}\rt)W_2(\cL(\Th^\lbd_n),\pi_\beta).
    \]
\end{proof}

\begin{lemma}\label{CV2}
	Let Assumption \ref{A1} hold. Then, for all $\beta \geq \max\{4/\gm,1/M\}$ with $M := 3\sqrt{L}/2+\sqrt{m}+\sqrt{2bL/\gm}$, the following bound holds:
	\begin{align*}
     &\E[V(\Th_\infty)]-\inf_{\Th \in \R^{N+1}} V(\Th) \\
      &\leq \frac{N+1}{2\beta}
      \log\left(
      \frac{4\pi e(b+(N+1)/\beta)\beta^2M^2}{\gm (N+1)}
      \right)
      +\frac{2}{\beta}
      -
      \frac{1}{\beta}\log\lf(\frac{2\pi^{(N+1)/2}}{(N+1)\Gamma((N+1)/2)}\rt).
      \end{align*}
\end{lemma}
\begin{proof}
 Let $\Th^*\in \arg\min_{\Th \in \R^{N+1}} V(\Th)$. According to \cite[Section 3.5]{raginsky2017}, we obtain, for all $\beta \geq 4/\gm$, that
\begin{equation}\label{CV21}
		\mathbb{E}[V(\Th_\infty)]
		\le
		\frac{N+1}{2\beta}
		\log\left(
		\frac{4\pi e(b+(N+1)/\beta)}{\gm (N+1)}
		\right)
		-
		\frac{1}{\beta}\log \Lambda,
	\end{equation}
	where
	\[
	\Lambda:=\int_{\mathbb{R}^{N+1}} e^{-\beta V(x)}\,dx
	\]
	is the normalization constant. Then,
	\begin{equation}\label{CV22}
		\log \Lambda
		=
		\log\int_{\mathbb{R}^{N+1}} e^{-\beta V(x)}\,dx
		=
		-\beta V(\Th^*)
		+
		\log\int_{\mathbb{R}^{N+1}}
		e^{\beta(V(\Th^*)-V(x))}\,dx .
	\end{equation}
Since $h(\Th^*) = \nabla V(\Th^*) = 0$, Lemma \ref{dissipativity} implies
$|\Th^*|\le \sqrt{2b/\gm}$. Using Lemma \ref{A2}, we have, for any $x\in\R^{N+1}$, that
	\[
	\begin{aligned}
		-\bigl(V(\Th^*)-V(x)\bigr)
		\le |V(\Th^*)-V(x)| &\le
		\int_0^1
		\left|
		\left\langle
		h\bigl(x+t(\Th^*-x)\bigr),\Th^*-x
		\right\rangle
		\right|\rd t \\
		&\le
		\int_0^1
		\left|h\bigl(x+t(\Th^*-x)\bigr)\right|
		|\Th^*-x|\rd t \\
		&\le
		\int_0^1
		\bigl(\sqrt{m}+\sqrt{L}|x|+t\sqrt{L}|\Th^*-x|\bigr)
		|\Th^*-x|\rd t \\
		&\le
		\int_0^1
		\bigl(\sqrt{m}+\sqrt{L}|\Th^*-x|+\sqrt{L}|\Th^*|
		+t\sqrt{L}|\Th^*-x|\bigr)
		|\Th^*-x|\rd t \\
		&\le
		\frac{3\sqrt{L}}{2}|\Th^*-x|^2
		+
		(\sqrt{m}+\sqrt{2bL/\gm})|\Th^*-x|.
	\end{aligned}
	\]
Therefore, with $M = 3\sqrt{L}/2+\sqrt{m}+\sqrt{2bL/\gm}$, we have
	\begin{equation*}
			\int_{\mathbb{R}^{N+1}} e^{\beta(V(\Th^*)-V(x))} \rd x\ge
			\int_{\mathbb{R}^{N+1}}
			e^{-\beta M (|\Th^*-x|+|\Th^*-x|^2)}
			\rd x,
	\end{equation*}
Using polar coordinates centered at $\Th^*$ yields
	\[
	\int_{\mathbb{R}^{N+1}}
	e^{-\beta M (|\Th^*-x|+|\Th^*-x|^2)}
	\rd x
	=
	\int_{
    \mathbb{S}_N
    }
	\int_0^\infty
	e^{-\beta M(r+r^2)}r^{N} \rd r \, \rd\omega
	=\frac{2\pi^{(N+1)/2}}{\Gamma(N+1)/2}
	\int_0^\infty
	e^{-\beta M(r+r^2)}r^{N} \rd r,
	\]
	where $\mathbb{S}_{N}$ is the surface area of the unit sphere in $\R^{N+1}$. Since $\beta M\ge 1$ by assumption,
	\[
	\begin{aligned}
			\int_{\mathbb{R}^{N+1}} e^{\beta(V(\Th^*)-V(x))} \rd x &\ge 	\frac{2\pi^{(N+1)/2}}{\Gamma((N+1)/2)}
		\int_0^\infty
		e^{-\beta M(r+r^2)}r^{N} \rd r\\
		&\ge
			\frac{2\pi^{(N+1)/2}}{\Gamma((N+1)/2)}
		\int_0^{(\beta M)^{-1}}
		e^{-\beta M(r+r^2)}r^{N}\rd r \\
		&\ge
			\frac{2\pi^{(N+1)/2}}{\Gamma((N+1)/2)}
		\int_0^{(\beta M)^{-1}}
		e^{-2}r^{N}\rd r \\
		&=
			\frac{2\pi^{(N+1)/2}}{(N+1)\Gamma((N+1)/2)} e^{-2}(\beta M)^{-(N+1)}.
	\end{aligned}
	\]
	Combining this bound with \eqref{CV22}, we get
	\[
	-\frac{1}{\beta}\log \Lambda
	\le
	V(\Th^*)
	+
	\frac{2}{\beta}
	-
	\frac{1}{\beta}\log\lf(\frac{2\pi^{(N+1)/2}}{(N+1)\Gamma((N+1)/2)}\rt)
	+
	\frac{N+1}{\beta}\log(\beta M).
	\]
	Substituting this estimate into \eqref{CV21} gives
	\begin{align*}
		&\mathbb{E}[V(\Th_\infty)]-V(\Th^*)\\
		&\le
		\frac{N+1}{2\beta}
		\log\left(
		\frac{4\pi e(b+(N+1)/\beta)}{\gm (N+1)}
		\right)
		+\frac{2}{\beta}
		-
		\frac{1}{\beta}\log\lf(\frac{2\pi^{(N+1)/2}}{(N+1)\Gamma((N+1)/2)}\rt)
		+
		\frac{N+1}{\beta}\log(\beta M)\\
		&=
		\frac{N+1}{2\beta}
		\log\left(
		\frac{4\pi e(b+(N+1)/\beta)\beta^2M^2}{\gm (N+1)}
		\right)
		+\frac{2}{\beta}
		-
		\frac{1}{\beta}\log\lf(\frac{2\pi^{(N+1)/2}}{(N+1)\Gamma((N+1)/2)}\rt).
	\end{align*}
	Since $V(\Th^*)=\inf_{\Th\in\R^{N+1}}V(\Th)$ by definition, the result follows.
\end{proof}

\section{Proofs of main results}\label{Proofs of main results}
\begin{proof}[\textbf{Proof of Theorem \ref{thm1}}]
Fix $\lbd \in (0,\lbd_0)$. Let $\cL(\Th_0)=\nu$ and denote by $(P_t)_{t\geq0}$ the Markov semigroup of the SDE \eqref{SDE} and by $\{\bar{P_n}\}_{n\in\N_0}$ the discrete-time semigroup of the SGLD algorithm $(\Th_n^{\lbd})_{n\in \N_0}$. Let $n\geq n_0$ with $n_0$ defined in \eqref{n0}. Then, the following holds by the triangle inequality
\begin{align}\label{thm1i}
    W_2(\cL(\Th^{\lbd}_n),\pi_\beta) &\leq W_2(\nu \bar{P}_n, \nu \bar{P}_{n-n_0}P_{n_0\lbd})+  W_2(\nu \bar{P}_{n-n_0}P_{n_0\lbd},\pi_\beta) \nonumber\\
    &\leq W_2(\nu \bar{P}_{n-n_0} \bar{P}_{n_0}, \nu \bar{P}_{n-n_0}P_{n_0\lbd})+  W_2(\nu \bar{P}_{n-n_0}P_{n_0\lbd},\pi_\beta)\nonumber\\
&=W_2(\cL(\vartheta^{\lbd,n}_{n_0}),\cL(\bar{\zeta}^{\lbd,n}_{n_0}))+W_2(\cL(\bar{\zeta}^{\lbd,n}_{n_0}),\pi_\beta).
\end{align}
The first term is controlled by Lemma~\ref{C5} with $t=n_0$,
\begin{equation}\label{thm1ii}
    W_2(\cL(\vartheta^{\lbd,n}_{n_0}),\cL(\bar{\zeta}^{\lbd,n}_{n_0})) \leq C_5 \lbd^{1/4}.
\end{equation}
Moreover, applying Proposition \ref{prop2} with initial law $\cL(\Th^{\lbd}_{n-n_0})$ over the time horizon $n_0 \lbd$ yields, for all $n\geq n_0$, that
\begin{equation}\label{thm1iii}
    W_2(\cL(\bar{\zeta}_{n_0}^{\lbd,n}),\pi_\beta) \leq C_{w}e^{-C_r n_0 \lbd} W_2(\cL(\Th^{\lbd}_{n-n_0}),\pi_\beta) \leq e^{-1}W_2(\cL(\Th^{\lbd}_{n-n_0}),\pi_\beta),
\end{equation}
where the last inequality holds by \eqref{n0}.
Combining \eqref{thm1i}--\eqref{thm1iii} gives, for all $n \geq n_0$, that
\[
 W_2(\cL(\Th^{\lbd}_n),\pi_\beta) \leq C_5 \lbd^{1/4} + e^{-1}W_2(\cL(\Th^{\lbd}_{n-n_0}),\pi_\beta).
\]
Applying Lemma \ref{recurssion} with $\varepsilon = C_5 \lbd^{1/4}$ and $\varrho = e^{-1}$ gives, for all $n\in \N_0$,
\[
 W_2(\cL(\Th^{\lbd}_n),\pi_\beta) \leq \frac{ C_5 \lbd^{1/4} }{ 1-e^{-1}}+e^{-(n/n_0-1)} \sup_{\mathfrak{n}\in\{0,...,n_0-1\}} W_2(\cL(\Th^{\lbd}_\mathfrak{n},\pi_\beta)),
\]
where the RHS of the last two inequalities are finite due to Lemma \ref{C6}. Moreover, the following inequality holds by \eqref{n0},
\[
\frac{n}{n_0} \geq \frac{n}{(\ln C_w+1)/(C_r \lbd)+1}\geq \frac{C_r n \lbd}{\ln C_w +1 +C_r \lbd_0} = \frac{n \lbd}{T_0}.
\]
This, Lemma \ref{C6}, and the fact that $1/(1-e^{-1}) \leq 2$ yield that for all $n \in \N_0$,
\[
W_2(\cL(\Th^{\lbd}_n),\pi_\beta) \leq 2C_5 \lbd^{1/4} +eC_6 e^{-n\lbd/T_0}.
\]
\end{proof}
\raggedbottom
\begin{proof}[\textbf{Proof of Theorem \ref{thm2}}]
Fix $n \in \N_0$, $\lbd \in (0,\lbd_0)$, and $\beta \geq \max\{4/\gm,1/M\}$. By Lemmas \ref{CV1} and \ref{CV2},
\begin{align*}
\E[V(\Th^\lbd_n)]-\inf_{\Th \in \R^{N+1}}V(\Th)&= \big(\E[V(\Th^\lbd_n)]-\E[V(\Th_\infty)]\big)+\big(\E[V(\Th_\infty)]-\inf_{\Th \in \R^{N+1}}V(\Th)\big)\\
&\leq  C_{V}W_2(\cL(\Th^{\lbd}_n), \pi_\beta)+	\frac{N+1}{2\beta}
\log\left(
\frac{4\pi e(b+(N+1)/\beta)\beta^2 M^2}{\gm (N+1)}
\right)\\
&\quad +\frac{2}{\beta}
-
\frac{1}{\beta}\log\lf(\frac{2\pi^{(N+1)/2}}{(N+1)\Gamma((N+1)/2)}\rt).
\end{align*}
Substituting the estimate from Theorem~\ref{thm1} for
$W_2\bigl(\cL(\Th_n^\lbd),\pi_\beta\bigr)$ completes the proof.
\end{proof}

\section{Supporting Lemma}
\begin{lemma}\cite[Lemma 14]{lytras2026tamed}\label{recurssion}
Let $n_0 \in \N$, $\varepsilon >0$, and $\varrho \in(0,1).$ If a non-negative sequence $\{a_n\}_{n\in\N_0}$ satisfies
\[
a_n \leq \varepsilon +\varrho a_{n-n_0}, \quad \forall n\geq n_0,
\]
then
\[
a_n \leq \frac{\varepsilon}{1-\varrho}+\max_{i\in\{0,...,n_0-1\}}a_i \varrho^{n/n_0-1}, \quad \forall n \in \N_0.
\]    
\end{lemma}
\begin{proof}
Write $n=jn_0+i$ with $j=\lfloor n/n_0 \rfloor$ and $i \in \{0,...,n_0-1\}$. Iterating the hypothesis $j$ times gives that
\[
a_n \leq \varepsilon(1+\varrho+\cdots+\varrho^{j-1})+\varrho^{j}\max_{i\in\{0,...,n_0-1\}}a_i\leq \frac{\varepsilon}{1-\varrho}+\varrho^{j}\max_{i\in\{0,...,n_0-1\}}a_i.
\]
Moreover, $\varrho^j \leq \varrho^{n/n_0 -1}$ for $\varrho \in (0,1)$ since $j \geq n/n_0 -1$.
\end{proof}

\begin{table}[H]
\centering
\caption{Analytic expressions of constants}
\label{table1}

{\footnotesize
\renewcommand{\arraystretch}{1.25}
\begin{tabularx}{\textwidth}[H]
{ 
	>{\hsize=0.45\hsize\linewidth=\hsize}X
	>{\hsize=0.2\hsize\linewidth=\hsize}X|
	>{\hsize=0.45\hsize\linewidth=\hsize}X|
	>{\hsize=2.9\hsize\linewidth=\hsize}X
} 
\hline
\hline
{CONSTANT} & & DIMENSION& FULL EXPRESSION\\
\hline
Lemma \ref{A2}	& $m$	&  --- 			&$2(1+1/(1-\bar{q}))^2+2\E[\mathcal{M}(X_0)^2]/(1-\bar{q})^2$\\
&$L$ &  --- 			&$8\gm^2$\\
\hline
Lemma \ref{A3}			& $r$  				&--- 					&$ 2(1+\E[\mathcal{M}(X_0)])/(\gm(1-\bar{q}))$ \\
\hline
Lemma \ref{dissipativity}			& $b$  				&  ---  					&$b=\max\left\{
		m/(2\gm),\;
		\sqrt{m}r+\left(\sqrt{L}+\gm/2\right)r^2
		\right\}$ \\
\hline	
Lemma \ref{A4}						& $k$  				&---  	
&$\lf(\E[\mathcal{M}(X_0)]/(1-\bar{q})-2\gm\rt)_+$ \\
\hline	
Proposition \ref{prop2}			
& $\rho(\beta)$  & --- 						&$ 2/(\beta \gm) e^{\beta (4r(\sqrt{m}+\sqrt{L}r)+\gm r^2/2)}$ \\
& $C_{w}$  & --- 	&$\sqrt{e\big(1+\beta k \rho(\beta)/2\big)}$ 
	 \\
& $C_{r}$  & --- 	&$2/(\beta \rho(\beta))$ \\
\hline
Remark \ref{remarkn0}					& $T_0$  				&---  	
&$(\ln C_w +1)/C_r+\lbd_0 $ \\
\hline	
Lemma \ref{C1}					& $C_1$  				&$\mathcal{O}(N)$				&$(4/\gm)(b+(N+1)/\beta)$ \\
\hline	
Lemma \ref{C2}					& $C_2$  				&$\mathcal{O}(N)$				&$(2b+\gm m/L + 2(N+1)/\beta)/(\gm - \lbd L)$ \\
\hline	
Lemma \ref{C3}					& $C_3$  				&$\mathcal{O}(N)$				&$2\gm C_2  + 2\gm m/L +4(N+1)/\beta$ \\
\hline	
Lemma \ref{C4}					& $C_4$  				&$\mathcal{O}(N)$				&$C_2+2 T_0\lf((N+1)/\beta +b\rt)$ \\
\hline	
Lemma \ref{C5}					& $C_5$  				&$\mathcal{O}(\sqrt{N})$				&$\begin{aligned}
			&\sqrt{2T_0}e^{2kT_0}
			\big(2\gm\mathbb{E}|\Th_0|^2+C_3
			\big)^{1/4}  \\
			&\hspace{-2mm}\times
			\big(
			2k
			\left(2\gm\mathbb{E}|\Th_0|^2+C_3
			\right)^{1/2}+
			\big(
			4m
			+4L\mathbb{E}|\Th_0|^2
			+2L(C_2+C_4)
			\big)^{1/2}
			\big)^{1/2}
		\end{aligned}$ \\
\hline	
Lemma \ref{C6} 		& $C_6$ 				&$\mathcal{O}(\sqrt{N})$		& $\sqrt{2(\E|\Th_0|^2+C_2) +2(\gm+2b+2(N+1)/\beta)/\gm}$\\
\hline
Lemma \ref{CV1} 		& $C_V$ 				&$\mathcal{O}(\sqrt{N})$		& $\sqrt{m}+\sqrt{L}\sqrt{\E|\Th_0|^2+C_2}/2+\sqrt{L}\sqrt{(\gm+2b+2(N+1)/\beta)/\gm}/2$\\
\hline
\hline
\end{tabularx}
}
\end{table}
\end{document}